\documentclass[a4paper,fleqn]{cas-dc}
\usepackage[numbers]{natbib}
\usepackage[utf8]{inputenc}
\usepackage{amsmath,amsfonts}
\usepackage{graphicx}
\usepackage{textcomp}
\usepackage[titles]{tocloft}
\usepackage[listings,skins]{tcolorbox}
\usepackage{breakurl}
\usepackage{hyperref}  
\usepackage{algpseudocode}
\usepackage[font=footnotesize,labelfont=bf]{caption}\usepackage{subcaption}
\usepackage{xspace}
\usepackage{pbalance}
\usepackage[linesnumbered]{algorithm2e}
\usepackage{adjustbox}
\usepackage{multirow}
\usepackage{tikz}
\usepackage{listings}

\newcommand{\etal}{et al.\@\xspace}

\newcommand{\wasm}{WebAssembly\xspace}

\definecolor{maroon}{cmyk}{0,0.87,0.68,0.32}
\definecolor{commentgreen}{RGB}{176, 176, 176}
\definecolor{greenplot}{RGB}{192,224,194}
\definecolor{orangeplot}{RGB}{ 249,217, 186}
\definecolor{eminence}{RGB}{108,48,130}
\definecolor{weborange}{RGB}{249,217,186}
\definecolor{frenchplum}{RGB}{129,20,82}
\definecolor{darkgreen}{RGB}{10, 92, 10}
\definecolor{celadon}{rgb}{0.67, 0.88, 0.69}

\newcommand{\thevariations}{6\ }

\usepackage{tabularx}

\usepackage{float}
\floatstyle{plaintop}
\restylefloat{table}
\restylefloat{figure}

\PassOptionsToPackage{hyphens}{url}\usepackage{hyperref}

\newcommand{\live}{%
    \begin{tikzpicture}%
    \fill [darkgreen] (0.0,0.0) circle [radius=0.07cm];
    \end{tikzpicture}%
}

\makeatletter

\newenvironment{btHighlight}[1][]
{\begingroup\tikzset{bt@Highlight@par/.style={#1}}\begin{lrbox}{\@tempboxa}}
{\end{lrbox}\bt@HL@box[bt@Highlight@par]{\@tempboxa}\endgroup}

\definecolor{commentgreen}{RGB}{176, 176, 176}
\definecolor{rowcolor}{cmyk}{0,0.87,0.68,0.32}
\definecolor{rowcolor2}{cmyk}{ 20, 0, 37, 34}

\definecolor{eminence}{RGB}{108,48,130}
\definecolor{weborange}{RGB}{255,165,0}
\definecolor{frenchplum}{RGB}{129,20,82}
\definecolor{darkgreen}{RGB}{10, 92, 10}

\definecolor{celadon}{rgb}{0.67, 0.88, 0.69}

\newcommand\btHL[1][]{%
  \begin{btHighlight}[#1]\bgroup\aftergroup\bt@HL@endenv%
}
\def\bt@HL@endenv{%
  \end{btHighlight}%   
  \egroup
}
\newcommand{\bt@HL@box}[2][]{%
  \tikz[#1]{%
    \pgfpathrectangle{\pgfpoint{1pt}{0pt}}{\pgfpoint{\wd #2}{\ht #2}}%
    \pgfusepath{use as bounding box}%
    \node[anchor=base west, fill=orange!30,outer sep=0pt,inner xsep=1pt, inner ysep=0pt, rounded corners=3pt, minimum height=\ht\strutbox+1pt,#1]{\raisebox{1pt}{\strut}\strut\usebox{#2}};
  }%
}
\makeatother

\newtheorem{metric}{\textbf{Metric}}

\newtheorem{defi}{\textbf{Definition}}

\newlistof{todo}{td}{List of TODOs}

\newcommand{\rev}[1]{#1}

\newcommand{\repourl}{\url{https://github.com/ASSERT-KTH/wasm_evasion}}

\begin{document}\sloppy

\hyphenation{Web-Assembly}
\hyphenation{Java-Script}

%\let\WriteBookmarks\relax
%\def\floatpagepagefraction{1}
%\def\textpagefraction{.001}

% Define the WAT language
\lstdefinelanguage{WAT}{
    otherkeywords={},
    morekeywords=[1]{i32,f32,i64,f64},
    morekeywords=[2]{0},
    morekeywords=[3]{add,const,mul,shl,get,rem_s,rem_u,ne,tee,sub,set,store},
    morekeywords=[4]{},
    morekeywords=[5]{global, get_global, mut, set_global, export, import,loop, memory, data, get_local,if, block,module, then,set_local,call,br_if,end, br,all,call_indirect,local,global,module, func, @custom, param, result, type},
    morekeywords=[6]{=,;},
    morekeywords=[7]{(,),[,],.},
    sensitive=false,
    morecomment=[l]{;},
    morecomment=[s]{;}{;},
    keywordstyle=[1]\color{eminence}\bfseries,
    keywordstyle=[3]\color{frenchplum},
    keywordstyle=[5]\color{darkgreen}\bfseries,
    commentstyle=\color{commentgreen},
    stringstyle=\color{darkgreen}
}
\lstdefinestyle{nccode}{
        numbers=none,
        firstnumber=2,
        stepnumber=1,
        numbersep=10pt,
        tabsize=4, 
        showspaces=false,
        breaklines=true, 
        showstringspaces=false,
    moredelim=**[is][\btHL]{`}{`},
    moredelim=**[is][{\btHL[fill=black!10]}]{`}{`},
    moredelim=**[is][{\btHL[fill=celadon!40]}]{!}{!}
}

\lstdefinestyle{WATStyle}{
  numbers=left,
  stepnumber=1,
  numbersep=10pt,
  tabsize=4,
  showspaces=false,
  showstringspaces=true,
}

\newcommand\mytitle{WebAssembly Diversification for Malware Evasion}
\shorttitle{}
\shortauthors{Cabrera-Arteaga et~al.}

\title [mode = title]{\mytitle}

\author[kth-address]{Javier Cabrera-Arteaga}[orcid=0000-0001-9399-8647]\cormark[1]\ead{javierca@kth.se}
\author[kth-address]{Martin Monperrus}[orcid=0000-0003-3505-3383]\ead{monperrus@kth.se}
\author[kth-address]{Tim Toady,}[orcid=0000-0002-0209-2805]\ead{toady@eecs.kth.se}
\author[kth-address]{Benoit Baudry}[orcid=0000-0002-4015-4640]\ead{baudry@kth.se}
\cortext[cor1]{Corresponding authors}

\address[kth-address]{KTH Royal Institute of Technology, Stockholm, Sweden}

%%
%% The abstract is a short summary of the work to be presented in the
%% article.
\begin{abstract}
\revision{
  %\todo{R1 - state the number of used malware samples and the evasion accuracy, so readers can have a better notion of the proposal potential. }
  
  WebAssembly has become a crucial part of the modern web, offering a faster alternative to JavaScript in browsers. While boosting rich applications in browser, this technology is also very efficient to develop cryptojacking malware. This has triggered the development of several methods to detect cryptojacking malware. 
  However, these defenses have not considered the possibility of attackers using evasion techniques. 
  This paper explores how automatic binary diversification can support the evasion of WebAssembly cryptojacking detectors. We experiment with  a  dataset of 33 WebAssembly cryptojacking binaries and evaluate our evasion technique against two malware detectors: VirusTotal, a general-purpose detector, and MINOS, a WebAssembly-specific detector. 
  Our results demonstrate that our technique can automatically generate variants of WebAssembly cryptojacking that evade the detectors in 90\% of cases for VirusTotal and 100\% for MINOS. 
  Our results emphasize the importance of meta-antiviruses and diverse detection techniques and provide new insights into which WebAssembly code transformations are best suited for malware evasion. 
  We also show that the variants introduce limited performance overhead, making binary diversification an effective technique for evasion.

}
\end{abstract}

%%
%% Keywords. The author(s) should pick words that accurately describe
%% the work being presented. Separate the keywords with commas.
% \keywords{WebAssembly, diversification}
%% A "teaser" image appears between the author and affiliation
%% information and the body of the document, and typically spans the
%% page.

%%
%% This command processes the author and affiliation and title
%% information and builds the first part of the formatted document.
\maketitle

\section{Introduction}

%\todo{R1 - 2. Still referring to AVs, sometimes the paper sounds as making generic claims about AVs, without pointing to the specific AV components (https://www.sciencedirect.com/science/article/pii/S0167404821003242). Please be more specific about AV components, detection engines, and so on.}

%\todo{R1 - - It is not clear the motivation for the design of the proposed solution. Why binary diversification and not other technique? Authors should inform the reader what features the detectors are currently detecting and why diversifying it evades detection.}

% Into to Wasm and current state of Wasm cryptojacking
WebAssembly is a binary format that has become an essential component of the web. 
It first appeared in 2017 as a fast and safe complement for JavaScript \cite{haas2017bringing}.
The language provides low-level constructs enabling efficient execution, much closer to native code than JavaScript.
Since its inception, the adoption of WebAssembly has grown exponentially, even outside the web \cite{MEWE}.
Its early adoption by malicious actors is further evidence of WebAssembly's success.

The primary black-hat usage of WebAssembly is cryptojacking \cite{10.1145/3488932.3517411}. Such WebAssembly code mines cryptocurrencies on users' browsers for the benefit of malicious actors and without the consent of the users \cite{ 10.1145/3339252.3339261}.
The main reason for this phenomenon is that the core foundation of cryptojacking is: the faster, the better.
In this context, WebAssembly, a binary instruction format designed to be portable and fast, is a feasible technology for implementing and distributing cryptojacking over the web.
A Kaspersky report about the state of cryptojacking in the first three quarters of 2022 confirms the steady growth in the usage of cryptominers \cite{kasperksy}.
The report shows that Monero \cite{monero} is the most used cryptocurrency for cryptomining in the browser.
Attackers might hide WebAssembly cryptominers \cite{9566204} in multiple locations inside web applications.

%\todo{TBD: Add Botacin paper and mention Kaspersky extension for browsers.}
Antivirus and browsers provide support for detecting cryptojacking. 
For example, the Firefox browser supports the detection of cryptomining by using deny lists \cite{firefoxcrypto}.
The academic community also provides related work on detecting or preventing WebAssembly cryptojacking \cite{SEISMIC, minos, coinspy, MineThrotle, outgard,9286112}.
Yet, it is known that black-hats can use evasion techniques to bypass detection.
Only the previous work of Bhansali \etal \cite{10.1145/3507657.3528560} investigates the possibility of WebAssembly cryptojacking to evade detection techniques. This is a crucial motivation for our work, one of the first to study WebAssembly malware evasion.

Our work is different from Bhansali \etal's  \cite{10.1145/3507657.3528560} in the following aspects. 
First, we extend the evaluation of MINOS by using VirusTotal and, we empirically demonstrate that this latter is a valid cryptojacking malware meta-detector for WebAssembly  to be used as baseline (see \autoref{method}). 
Second, we conduct an evaluation of the correctness and efficiency of the Wasm variants, which provides insights into the trade-offs and limitations of  bytecode-level transformations for malware evasion in WebAssembly.
Our approach, performs bytecode transformations at the WebAssembly level, instead of source code based transformations like the Bhansali's technique. %\todo{XXX in the case of Bhansali's techniques}.
We focus on software diversification techniques in the spirit of Cohen  \cite{cohen1993operating}, as this technique has the ability to generate many variants, while the impact on the binary size and performance can be controlled through the selection of specific diversification transformations \cite{lundquist2016searching}.

In this paper, we design, implement and evaluate a  full-fledged evasion pipeline for WebAssembly.
Concretely, we use wasm-mutate as a diversifier \cite{wasm-mutate}, which implements 135 possible bytecode transformations, grouped into three categories: peephole operator, module structure, and control flow.
We demonstrate the effectiveness of our evasion technique against two cryptojacking detectors: VirusTotal, a general detection tool that comprises 60 antiviruses and, MINOS \cite{minos}, a WebAssembly-specific detector.

We evaluate our proposed evasion technique on 33 cryptojacking malware that we curated from the 8643 binaries of the wasmbench dataset \cite{Hilbig2021AnES}, to our knowledge, the most exhaustive collection of real-world WebAssembly binaries.
We experiment with the 33 binaries marked as potentially dangerous by at least one antivirus vendor of VirusTotal.
We empirically demonstrate that evasion is possible for all of these 33 real-world WebAssembly cryptojacking malware while using a WebAssembly-specific detector. 
Remarkably, we find 30 cryptominers for which our technique successfully generates variants that evade VirusTotal.
Our set of malware includes 6 cryptojacking programs that are fully reproducible in a controlled environment. 
With them, we assess that our evasion method does not affect malware correctness and generates fully functional malware variants with minimal overhead.

Our work provides evidence that the malware detection community has opportunities to strengthen the automatic detection of cryptojacking WebAssembly malware. 
The results of this work are actionable, as we provide quantitative evidence on specific malware transformations on which detection methods can focus.
To sum up, the contributions of this work are:

\begin{itemize}
    \item A full-fledged cryptojacking malware evasion pipeline for WebAssembly, based on a state-of-the-art binary diversification. We provide the repository of the tool at \repourl.
    \item A systematic evaluation of our cryptojacking evasion pipeline, including effectiveness, performance, and correctness. 
    \item Actionable evidence on which transformations are better for evading WebAssembly cryptojacking detectors, calling for future work from the academic and the industrial community alike.

    \item A reproducible comparison of VirusTotal with MINOS, a WebAssembly-specific detector, showing the relevance of VirusTotal as a valid and practical cryptojacking meta-detector.

\end{itemize}

This paper is structured as follows. In \autoref{background}, we introduce WebAssembly for cryptomining and the state-of-the-art on malware detection and evasion techniques and its limitations for WebAssembly cryptojacking.
In \autoref{attack-model-section}, we instantiate and explain malware evasion for WebAssembly cryptojacking in a real scenario. 
We follow with the technical description of our malware evasion algorithms in \autoref{arch}. 
We formulate our research questions in \autoref{method}, answering them in \autoref{results}.
We discuss our research in \autoref{discussion}, in order to help future research projects on similar topics.
We finalize with our conclusions \autoref{conclusions}. 

\section{Background \& Related Work}
\label{background}

In this section, we introduce WebAssembly.
Besides, we illustrate its usage for cryptojacking. 
Then, we discuss how WebAssembly cryptojacking can be detected, and the most common techniques used to evade such detection.

{
\subsection{WebAssembly}
%\todo{R1 - - Authors need to explain in more detail how Wasm binaries work. Since the authors will present a feature diversification approach, the features to be diversified should be introduced here. Authors presented a few details related to VM execution in 2.1, but authors should go deeper on binary format description, for instance.}

%\todo{Not good. Just a code example, instantiate in 3.1 diversifier. Do not enumerate the sections, more abstract description of the language. Loop, unroll transformation in the 3.1 section. No need to understand deep section insights to understand diversification. 1 example, one no more than 2 paragraphs.}

WebAssembly (Wasm) is a binary instruction set meant initially for the web. 
It was adopted as a standard language for the web by the W3C in 2017, building upon the work of Haas et al. \cite{haas2017bringing}. One of Wasm's primary advantages is that it defines its own Instruction Set Architecture (ISA), which is both straightforward and platform-independent. As a result, a Wasm binary can execute on virtually any platform, including web browsers and server-side environments. Since its introduction, all major web browsers have implemented support for WebAssembly, reporting to be only 10\% slower than machine code during runtime.

WebAssembly programs are compiled ahead-of-time from source languages such as C/C++, Rust, and Go, utilizing compilation pipelines like LLVM. This allows Wasm to benefit from ahead-of-time compiling optimizations, improving its performance. 
A Wasm binary is comprised of sections, which are consecutive sequences of bytes in the binary file. 
In contrast to the absolute order of sections in Windows Portable Programs, sections in Wasm binaries have a relative order between them.
Thus, Wasm can be considered a more flexible binary format.

\lstdefinestyle{watcode}{
  numbers=none,
  stepnumber=1,
  numbersep=10pt,
  tabsize=4,
  showspaces=false,
  breaklines=true, 
  showstringspaces=false,
  moredelim=**[is][{\btHL[fill=black!10]}]{`}{`},
  moredelim=**[is][{\btHL[fill=celadon!40]}]{!}{!}
}

{\captionsetup{width=0.39\linewidth}
\noindent\begin{minipage}[t]{0.4\linewidth}
  \lstset{
    language=C,
    basicstyle=\footnotesize\ttfamily,
    columns=fullflexible,
  postbreak=\mbox{\textcolor{red}{$\hookrightarrow$}\space},
    breaklines=true}
    \begin{lstlisting}[label=example:cprogram,caption={C program containing function declaration, loops, conditionals and memory access.},captionpos=t]{Name}
int a[100];
void cc1(){
    for(int i = 0; i < 100; i++){
        a[i] = i;
    }
}

    \end{lstlisting}
    
\end{minipage}\hfill
\noindent\begin{minipage}[t]{0.6\linewidth}
\captionsetup{width=0.89\linewidth}

\lstset{
  language=WAT,
  style=watcode,
  breaklines=true, 
  basicstyle=\footnotesize\ttfamily,
  %numberstyle=\footnotesize,
  numbersep=2.5pt,
  %firstnumber=1,
  escapeinside={(*@}{@*)},
  %numbers=left,
  %postbreak=\mbox{\textcolor{red}{$\hookrightarrow$}\space},
  }
    \begin{lstlisting}[label=example:wasmprogram,caption={WebAssembly code for C code in \autoref{example:cprogram}.}, captionpos=t]{Name}

    
(module
  (@custom "producer" "llvm.." )
  (func (;5;) (type 1)
    (loop  ;; label = @1
      (if  ;; label = @2
        (i32.eqz
          (i32.ge_s
            (i32.load
              (local.get 0))
            (i32.const 100)))
        (then (i32.store
            (i32.add
              (i32.shl
                (i32.load
                  (local.get 0))
                (i32.const 2))
              (i32.const 1024))
            (i32.load
              (local.get 0)))
          (i32.store
            (local.get 0)
            (i32.add
              (i32.load
                (local.get 0))
              (i32.const 1)))
          (br 1 (;@1;)))))
    )
  (memory (;0;) 256)
  (global (;1;) i32 (i32.const 0))
  (export "memory" (memory 0)))
    \end{lstlisting}
\end{minipage}
}

WebAssembly programs operate on a virtual stack that allows for only four data types: i32, i64, f32, and f64. 
These same data types are used to annotate the numeric operations in the WebAssembly code.
Additionally, a WebAssembly program might include several custom sections.
For example, binary producers such as compilers use it to store metadata.
A WebAssembly code also declares memories and globals, which are used to store, manipulate and share data during program execution, e.g. to share data with the host engine of the WebAssembly binary.

WebAssembly is designed with isolation as a primary consideration. For instance, a WebAssembly binary cannot access the memory of other binaries or interact directly with a browser's built-in API, such as the DOM or the network. Instead, communication with these features is constrained to functions imported from the host engine, ensuring a secure and safe Wasm environment.
Moreover, control flow in WebAssembly is managed through explicit labels and well-defined blocks, which means that jumps in the program can only occur inside blocks, unlike regular assembly code.

%\todo{this paragraph mixes presentation of the listing and presentation of the language}
   In \autoref{example:cprogram}, we provide an example of a C program that contains a function declaration, a loop, a loop conditional, and a memory access. When the C code is compiled to WebAssembly, it produces the code shown in \autoref{example:wasmprogram}. The stack operations are folded with parentheses.
   The module in the example contains the components described previously.

\vspace{5mm}
\subsection{Malware in WebAssembly}
\label{background:crypto}

The use cases of WebAssembly in browsers focuses on computation-intensive activities such as gaming or image processing. 
Also, malign actors have taken advantage of WebAssembly to carry out their activities, and cryptojacking is the most common usage observed so far \cite{ 10.1145/3339252.3339261, newkid}.
The reason for this is that cryptojacking involves executing vast amounts of hash functions, which requires significant computing resources. 
In comparison to JavaScript, WebAssembly is significantly faster at handling  these intense hashing operations in the browser \cite{haas2017bringing}. 
% As a plus, the readability of Wasm code is comparatively more difficult for humans, making it a natural effective tool for obfuscating malicious code. 

Web cryptojacking is often carried out by including a malicious JavaScript+WebAssembly payload, which then execute on the victim's browser without their knowledge  \cite{9566204}.
For example, websites that offer illegal download or adult sites, often include cryptojacking in their webpages to generate passive income.
Since cryptojacking is difficult to detect and remove, it can remain on a victim's computer for an extended period,  continuing to consume resources and to generate income for the attacker. 
This lucrative form of malware does need vulnerabilities or stealing credentials.

\subsection{Malware Detection}
\label{sec:detection}
%\todo{R1 -  In 2.2, authors cite some picked malware detection papers. I suggest authors citing a malware detection survey instead to increase the coverage.}

Malware detection determines if a binary is malicious or not.
This process can be based on static, dynamic, or hybrid analysis \cite{8949524}. 
In this section, we highlight works in the area of malware detection.
Static-based approaches analyze the source code or the binary to find malign patterns without executing them. The literature reports a range of techniques, from simple checksum checking to advanced machine learning methods, that have subsequently been adopted by commercial antiviruses \cite{BOTACIN_AVs, arms}.
% in WebAssembly+
In the context of WebAssembly, MineSweeper is a detection method based on static analysis \cite{Konoth2018} of WebAssembly.
Its detection strategy depends on the knowledge of the internals of CryptoNight, one popular library for cryptomining. 
In the same context, MINOS is a state-of-art static detection tool that converts WebAssembly binaries to vectors for malware detection  \cite{minos}.

MINOS is a practical approach for detecting malicious WebAssembly binaries. 
It works by converting the Wasm binary's bytestream into a 100x100 grayscale image, which is then fed into a Convolutional Neural Network (CNN). 
The CNN has learned patterns in the image to classify it as either benign or malicious. 
This approach is similar to image-based methods used in other areas \cite{7811100}, e.g., for detecting Windows malware \cite{kalash, RandomShield}.
We believe MINOS is the optimal static approach to detect WebAssembly malware due to its simplicity and practicality, such as being easily implemented as a browser extension.

%\todo{R1 - In 2.2., it is not clear which features MINOS uses to detect WAsm malware. This information is important for this paper, as it is the main topic.
%}

Dynamic analysis for malware detection is based on the execution of the malware code to identify potentially dangerous behaviors \cite{10.1145/2089125.2089126}. 
Usually, this is done by monitoring some functions, such as API calls.
For example, BLADE \cite{blade} is a  Windows kernel extension that aims to eliminate drive-by malware installations.
It wraps the filesystem for browser downloads for which user consent has been involved. It thwarts the ability of browser-based exploits to download and execute malicious content surreptitiously.
% Dynamic approaches for WebAssembly
SEISMIC and MineThrotle also perform dynamic analyses  \cite{SEISMIC, MineThrotle}
on WebAssembly binaries to profile instructions that are specific to cryptominers. 
For example, cryptominers overly execute  \texttt{XOR} instructions.  
SEISMIC and MineThrotle use machine learning approaches to classify the binary as benign or malign based on collecting runtime profiles.
On the same topic, MinerRay \cite{9286112} detects cryptojacking in WebAssembly binaries by analyzing their control flow graph at runtime, searching for structures that are characteristic of encryption algorithms commonly used for cryptojacking.
CoinSpy is another malware detector based on dynamic analysis \cite{coinspy}. 
It uses a convolutional neural network to analyse the computation, network, and memory information caused by cryptojackers running in client browsers.

Hybrid approaches use a mix of static and dynamic detection techniques. 
The main reason to use hybrid approaches is the impracticability of executing the whole program.
Thus, only pieces of code that can be quickly executed are dynamically analyzed.
For example, AppAudit embodies a novel dynamic analysis for Android applications that can simulate the execution of some parts of the program \cite{appaudit}. 
For WebAssembly, Outguard \cite{outgard} trains a  Support Vector Machine model with a combination of cryptomining function names obtained statically and dynamic information such as the number of web workers used in the web application that is analyzed.

It is possible to combine several independent detectors into a meta-antivirus.
Each detector embeds some heuristics that are good at detecting specific types of malware \cite{MOSER}. Hence, their combination can effectively detect a broader range of malware, e.g., using relationship analysis.
VirusTotal \cite{virustotal, virustotalgoogle} is a consolidated meta-antivirus. 
VirusTotal operates with 60 antivirus vendors to provide malware scanning. 
Through its API, a program can be labeled by 60 antiviruses.
This aggregation is used to determine if an asset under analysis is malicious, e.g., by voting.
Previous works used VirusTotal to assess detection efficiency, \cite{10.1145/3355369.3355585} because it is a proxy to evaluate state-of-art techniques in combination with commercial antiviruses.
In this work, we follow the same methodology, using VirusTotal to assess our technique's ability to evade cryptojacking detectors.

While concerns have been raised about the use of VirusTotal for some malware and file type families \cite{BOTACIN2020101859}, it can be considered for WebAssembly cryptojacking detection.
We empirically highlight later in this paper that VirusTotal is slightly better  than the WebAssembly-specific detector MINOS regarding the detection of cryptojacking malware.

%\todo{R1 -  It is hard to claim VT as a detection approach. It's more a benchmark tool than a detection one. VT has some known drawbacks for detection, such as delays and differences between the VT AVs and the actual AVs at the endpoint. See 
%}
%\todo{On the other hand, our experiments can be easyly and fast reproduced, meaning that there is no time for labeling change.}
% https://www.sciencedirect.com/science/article/pii/S0167404820301310

%\todo{R1 - the previous paragraph reads as a conclusion / transition to the next subsection. We should move the Virustotal paragraph before: what is the most appropriate location? or should it have its own subsection?}

\subsection{Malware Evasion}

%\todo{R1 -  Authors state that: "ML-based detection can be bypassed in so-called malware evasion attempts.". In fact, malware evasion is a generic name for the evasion of any type of detection, not only ML-based. Signature-based detectors also suffer malware evasion. I guess authors were referring to adversarial ML attacks, thus Adversarial Examples (AEs).}

%\todo{TBD: Saying that all of them use ML is a bad assumption. According to BOTACIN.}
% Transition

%As discussed in the previous subsection, most work on malware detection for WebAssembly uses machine learning approaches. The limitations of learning-based approaches for malware detection have been highlighted by Li and colleagues \cite{arms}. {\color{blue} In particular, ML-based detection can be bypassed by so-called  adversarial generation attacks}.

Malware evasion techniques aim at avoiding malware detection \cite{surveymalware} 
Potential attackers use a wide range of techniques to achieve evasion, such as genetic programming \cite{castro2019aimed}.
With time, the techniques to avoid detection have grown in complexity and sophistication \cite{Aghakhani2020WhenMI}.
For example, Chua \etal \cite{chua} proposed a framework to automatically obfuscate Android applications' source code using method overloading, opaque predicates, try-catch, and switch statement obfuscation, creating several versions of the same malware.
Also, machine learning approaches have been used to create evading malware \cite{2021arXiv211111487D}, based on a corpus of pre-existing malware \cite{Bostani2021EvadeDroidAP}.
While most approaches try to break static malware detectors, more sophisticated techniques avoid dynamic detection, usually involving throttling techniques or dynamic anti-analysis checks \cite{Lu2013WeaknessesID, payer2014embracing}. Wang proposes the concept of 
Accrued Malicious Magnitude (AMM) to identify which malware features should be manipulated to maximize the likelihood of evading detection   \cite{wang2021exposing}.

In the context of WebAssembly, malware evasion is nearly unexplored.
Only Romano \etal \cite{romano2022wobfuscator} recently proposed wobfuscator, a code obfuscation technique that transforms JavaScript code into a new JavaScript file and a set of WebAssembly binaries.
Their technique mostly focuses on JavaScript evasion and not WebAssembly evasion.

Bhansali \etal \cite{10.1145/3507657.3528560} propose a technique where WebAssembly binaries are transformed while maintaining the functionality with seven different source code obfuscation techniques. 
They evaluate the effectiveness of the techniques against MINOS \cite{minos}.
They show these transformations can generate malware variants that evade the MINOS classifier.

% Limitations

%\todo{R1 - - In 2.3, it is not clear what is missing on the (few) existing evasion generators.}

\section{WebAssembly cryptojacking malware evasion in practice}
\label{attack-model-section}
%\todo{R1 - 1. Add a Threat Model section. It is important to clarify to the reader what are the expected attacks, defenses, and context. For instance, authors need to clarify that they expect attacks to occur in the browser via deceptive links. Authors should also clarify what are the expected defenses. The use of AV works for the purpose of an oracle, but how it would look like in a real system? DO authors expect browsers to be maliciousness aware? Or Do authors expect AVs to have security plugins to detect browser-based threats?
%}

%\todo{before introducing the figure, we need a bit of context. For example, discuss the fact that we introduce a scenario and our goal is to demonstrate the feasibility of the attack (not to introduce a defense against this attack)}

\autoref{attack-model} illustrates our attack scenario:
a practical WebAssembly cryptojacking attack consists of three components: a WebAssembly binary, a JavaScript wrapper, and a backend cryptominer pool. 
The WebAssembly binary is responsible for executing the hash calculations, which consume significant computational resources. 
The JavaScript wrapper facilitates the communication between the WebAssembly binary and the cryptominer pool.
Overall, a successful cryptojacking attack on a victim's browser consists in the following sequence of steps. 
First, the victim visits a web page infected with the cryptojacking code. 
The web page establishes a channel to the cryptominer pool, which then assigns a hashing job to the infected browser. 
The WebAssembly cryptominer calculates thousands of hashes inside the browser, in parallel using multiple browser workers \cite{workers}. 
Once the malware server receives acceptable hashes, it is rewarded with cryptocurrencies for the mining. Then, the server assigns a new job, and the mining process starts over.

Some detection techniques discussed in \autoref{sec:detection} can be  deployed in the browser directly to prevent cryptojacking.
The primary objective of our work is to demonstrate the possibility of using code diversification to bypass cryptojacking defenses. 
Concretely, the following workflow can happen to successfully evade placed defenses:

i) The user visits a webpage that contains a cryptojacking malware, which utilizes network resources to execute, (1) and (2) in \autoref{attack-model}. Cryptojacking malware can be injected through malicious browser extensions, malvertising, compromised websites, or deceptive links \cite{9566204}. 
ii) A malware detector blocks WebAssembly binaries that are identified as malicious (3). The malware detector system can be implemented locally or remotely. For instance, a proxy can intercept and send network resources to an external detector through the detector's API. 
iii) The attacker, based on a malware oracle, crafts a WebAssembly cryptojacking malware variant that evade the detection (4). 
iv) The attacker delivers the modified binary instead of the original one (5), which initiates the cryptojacking process and compromises the browser (6).

\begin{figure}
    \centering
    \includegraphics[width=0.98\linewidth]{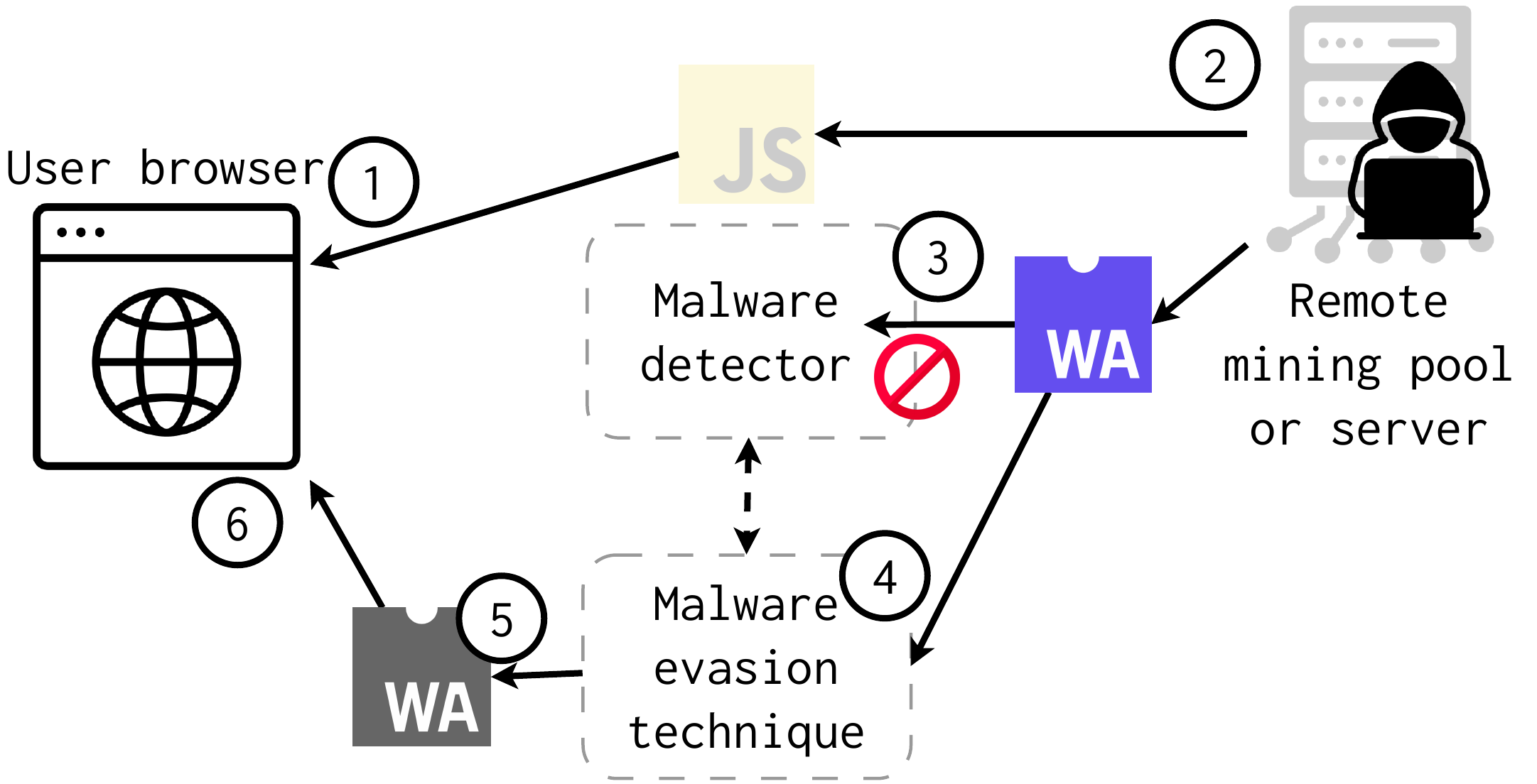}
    \caption{WebAssembly evasion in practice. The user visits a webpage containing cryptojacking malware that uses network resources to operate. A malware detector blocks identified malicious WebAssembly binaries. The attacker, using a malware oracle, creates a WebAssembly cryptojacking malware variant that evades detection. Finally, the attacker delivers the modified binary, initiating the cryptojacking process and compromising the browser.}
    \label{attack-model}
\end{figure}

%\todo{TBD J: maybe start with the attacker to be more like other threat model papers.}

%\todo{in general, never use bullet points in scientific papers, inline}

%\todo{the following paragraph aims at stressing the need for 'fast' diversification; let's make this intention more explicit}

The idea is that attackers rapidly diversify their WebAssembly code to stay ahead of the defense system and maintain successful cryptojacking operations. 
Crucially, attackers must ensure that the diversified binaries they use for cryptojacking meet specific performance requirements, which is an aspect we will study in \autoref{arch}. 
% This requirement is why we chose diversification over obfuscation.  Obfuscation tools can generate program variants that are slower and less performant than the original program \cite{HOSSEINZADEH201872}.  In contrast, WebAssembly diversifiers can be tuned to generate variants that span the full performance spectrum \cite{crow,MEWE}, from faster to slower.

\section{Diversification for Malware Evasion in WebAssembly}
\label{arch}

%\todo{R1 - Change Oracle to Malware Oracle. Change non-interesting by to benign. Change "Wa" to be explicitly annotated with cryptojacking. Fitness function instead of Feedback.}

In this section, we explain a technique for potential attackers to craft a WebAssembly binary that evades detection (steps 4, 5 and 6 in \autoref{attack-model}).

In \autoref{generic} we illustrate our generic architecture for the malware evasion component.
The workflow starts by passing a WebAssembly malware binary to a software diversifier (1).
The diversifier generates binary variants, which are passed to a malware oracle (2). The oracle returns labeling feedback for the binary variant: malware or benignware.
The oracle result is the input for a fitness function that steers the construction of a new binary on top of the previously diversified one.
This process is repeated until the malware oracle marks the mutated binary as benign or a timeout is reached (4).
For the sake of open science and for fostering research on this important topic, our implementation is made publicly available on GitHub: \repourl.

\begin{figure*}
    \centering
    \includegraphics[width=0.9\linewidth]{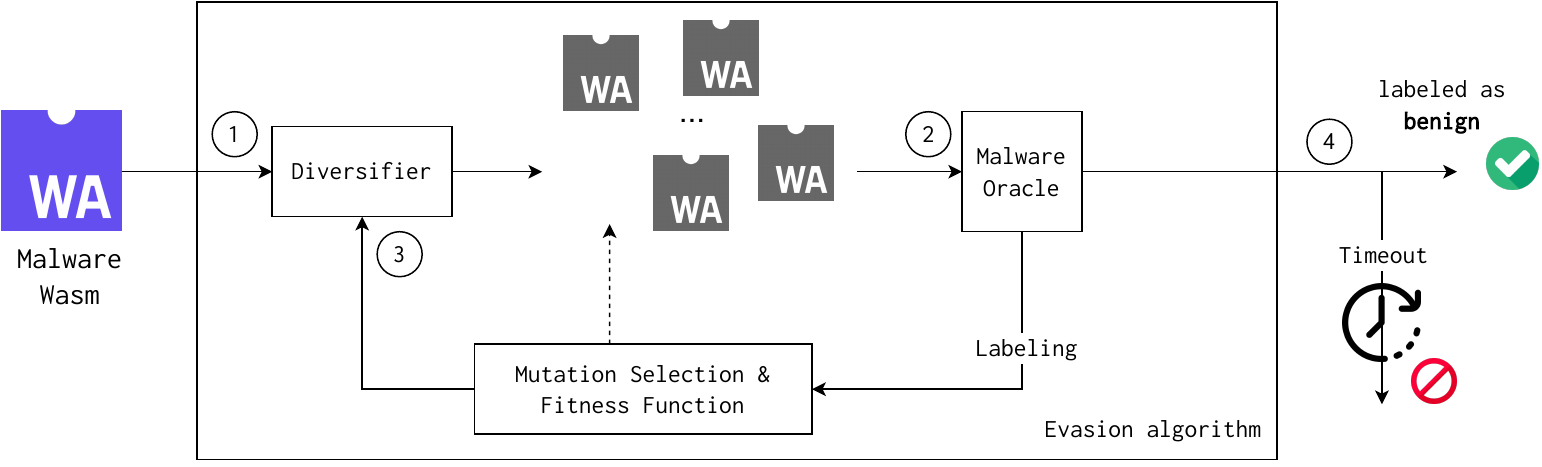}
    \caption{Our original workflow of binary diversification for malware evasion in WebAssembly. The workflow begins with a WebAssembly malware binary sent to a software diversifier. The diversifier creates binary variants, which are analyzed by a malware oracle that labels them as malware or benignware. The oracle's results guide the development of a new binary based on the previous one. This process continues until the malware oracle labels the mutated binary as benign or a timeout occurs.}
    \label{generic}
\end{figure*}

\subsection{Diversifier}
\label{diversifier}

Conceptually, our approach is parametrized by a semantic preserving diversifier \cite{cohen1993operating}}. For our prototype implementation, we select one diversifier that supports wasm-to-wasm diversification and performs almost non-costly transformations: wasm-mutate \cite{wasm-mutate}.
This tool takes a WebAssembly module as input and returns a set of variants of that module. 
wasm-mutate follows the notion of program equivalence modulo input \cite{moduloinputpaper}, i.e., the variants should provide the same output for the same program inputs. 
It represents the search space for new variants as an e-graph \cite{2020arXiv200403082W}, and it exploits the property that any traversal through the e-graph represents a semantically equivalent variant of the input program.
wasm-mutate can generate thousands of semantically equivalent variants from a single binary in a few minutes.

wasm-mutate defines 135 possible transformations on an input WebAssembly binary, grouped into three categories.
% Peephole driver
The peephole operator is the first category. It is responsible for rewriting instruction sequences in function bodies. It selects a random instruction within the binary functions and applies one or more of the 125 rewrite rules.
Finally, it re-encodes the WebAssembly module with the new, rewritten expression to produce a binary variant. 
% Module structure driver
The second category of transformations in wasm-mutate implements module structure transformations.
It operates at the level of the WebAssembly binary structure.
It includes the following eight transformations:  add a new type definition, add/modify custom sections, add a new function, add a new export, add a new import, add a new global, remove a type and remove a function.
The third transformation category is on the control flow graph of a function code level.
wasm-mutate performs two possible transformations: unroll a loop or swap the branches of a conditional block.

%\todo{TBD: Bintuner is not classical diversification paper :|}

%\todo{R1 - Authors should mention that diversification is typically used in desktop binaries, including malware. More specifically, authors should refer to the Bintuner classical paper: https://dl.acm.org/doi/10.1145/3453483.3454035}

%\todo{R1 - Whereas it is clear that authors used a Wasm-to-Wasm diversifier tool, it is not clear how project decisions were made. If such framework already exists, it is because diversification was previously used in Wasm. So, what was it used for before? What insight authors used to move from this other domain to malware evasion?}

%\todo{R1 - Since the related works in the Wasm domain are limited, authors could try to establish relations with diversification approaches from other domains, such as desktop. The presented transformations for Wasm, for instance, look similar to the ones of LLVM-obfuscator: https://github.com/obfuscator-llvm/obfuscator}

%\todo{ R1 -  Similarly, transformations involving sections and so on are similar to the ones used in typical Windows executables: https://arxiv.org/abs/2003.13526}

The decision of using wasm-mutate as a diversifier is based on three key factors. 
First, while diversification approaches for WebAssembly from LLVM sources exist \cite{crow, MEWE}, compiler-based diversification may include compiler fingerprints in the built binaries, which is bad for stealthy evasion. 
Second, while optimization-based approaches could be used to diversify WebAssembly binaries from source code \cite{BinTuner}, optimizations are usually all applied at once, providing a smaller diversification space, hence fewer opportunities for evasion.
Finally, wasm-mutate is a tool that implements many useful semantically equivalent transformations, making it well-suited as a diversifier with minimal engineering effort.
Therefore, using wasm-mutate as a diversifier provides a practical approach for evasion in WebAssembly. We believe that attackers would take the same path and reach the same conclusion.

\lstdefinestyle{watcode}{
  numbers=none,
  stepnumber=1,
  numbersep=10pt,
  tabsize=4,
  showspaces=false,
  breaklines=true, 
  showstringspaces=false,
  moredelim=**[is][{\btHL[fill=orange!40]}]{`}{`},
  moredelim=**[is][{\btHL[fill=celadon!40]}]{!}{!}
}

\lstset{
  language=WAT,
  caption={\wasm\ code  for \autoref{CExample}.},
  style=watcode,
  breaklines=true, 
  basicstyle=\footnotesize\ttfamily,
  %numberstyle=\footnotesize,
  numbersep=2.5pt,
  %firstnumber=1,
  escapeinside={(*@}{@*)},
  %numbers=left
  label=WASMExample3}
  
    \begin{lstlisting}[label=example:wasmprogrammutated,caption={Wasm-mutate transformation applied over \autoref{example:wasmprogram}.}, captionpos=t]{Name}


(module
 `- (@custom "producer" "llvm.." )`
  (func (;5;) (type 1)
    (loop  ;; label = @1
      (if  ;; label = @2
        (i32.eqz
          (i32.ge_s
            (i32.load
              (local.get 0))
            (i32.const 100)))
        (then (i32.store
            (i32.add
               !(i32.mul!
               !  (i32.load !
               !    (local.get 0)) !
               !  (i32.load !
               !    (local.get 0)) !
               )
              (i32.const 1024)
            )
            (i32.load
              (local.get 0)))
          (i32.store
            (local.get 0)
            (i32.add
              (i32.load
                (local.get 0))
              (i32.const 1)))
          (br 1 (;@1;)))))
    )
  ...
    \end{lstlisting}

For the sake of illustration, in \autoref{example:wasmprogrammutated}, we present a variant of the WebAssembly code shown in \autoref{example:wasmprogram}. 
We generate this variant using the wasm-mutate diversifier, with two transformations. 
The changes made to the original code are highlighted in orange and green.
The first transformation, highlighted in orange, involves removing the custom section that indicates the producer of the original binary. 
The second transformation, highlighted in green, replaces the shift-left operation in the original binary with two consecutive multiplications of the same value.

\subsection{Malware Oracle}
\label{malware_oracle}

%\todo{ R1 -  In 3.2. it is OK to use VT as oracle. However, there are corner cases to use it as a numerical oracle. Notice that the number of AVs reported by VT might change from query to query. How do authors handle this case? }

To determine whether a given sample is malicious or not, we rely on a malware oracle. 
The simplest form of malware oracle is a binary classifier that outputs a label of \{malware, benignware\}. 
In addition to binary classifiers, we also consider numerical oracles that provide a value representing the likelihood that a sample is malicious.

In our experiments, we use VirusTotal \cite{virustotal} as our malware oracle.
VirusTotal operates with  60 antivirus
vendors to provide malware scanning. 
Users submit binaries, and they receive labels from different vendors.
The resulting 60 labels can then be used to determine if a queried asset is malicious, e.g., by voting.
VirusTotal can be used as both a binary and a numerical oracle \cite{251586}.
We use VirusTotal as a binary oracle by returning malware if at least one vendor classifies the binary as malware.
We also use VirusTotal as a numerical oracle, using the number (between 0 and 60) of oracles labeling a binary as malware.

research \cite{BOTACIN2020101859}, classification labels assigned to samples by vendors in VirusTotal can change over time due to new antivirus releases. 
In our work, we operate under the assumptions outlined in \autoref{attack-model-section} and consider a scenario where an attacker develops and executes an evasion technique in under an hour. 
This timeframe is significantly shorter than the time it takes for classification labels to change in VirusTotal, which typically takes several days.

\subsection{Transformation Selection \& Fitness Function}
\label{sec:evasion-algorithms}

The third step of our workflow consists of two actions towards synthesizing a malware variant:  select a wasm-mutate transformation to apply; and determine if the transformation is applied.
This latter decision depends on: 1) the result of the malware oracle at the previous iteration and 2) an estimation of the ability of the new transformation to generate an evading binary. For this estimation, we implement two variations of our evasion algorithm.

First, we use a binary malware oracle. In this context, we always apply the transformation. This is the \emph{baseline evasion algorithm}, which is discussed in more detail later.

% Second usage
The second variation of the evasion algorithm includes a  fitness function that uses a numerical oracle to estimate if the transformation should be applied. 
%The numerical oracle returns a value between 0 and 60. 
The fitness function uses the information from VirusTotal and is the total number of vendors (between 0 and 60) that label a binary as malware:
\begin{equation*}
    FF(m)= \Sigma_{i=0}^{i=60} \begin{cases}
        1\text{ if }v_i(m)\text{ returns \texttt{malware}}\\
        0\text{ i.o.c}
    \end{cases}
\end{equation*}
This fitness function is used in our second proposed algorithm and is discussed in details below.

\subsubsection{Baseline Evasion}

The baseline evasion algorithm is described in \autoref{alg:naive}.
It uses VirusTotal as a binary oracle (true if at least one VirusTotal detects malware, false otherwise).
In this algorithm, step 1 is in line 3, steps 2 and 4 are in lines 4 to 6, and step  3 is in line 7.
Each iteration of this algorithm, ``stacks'' a transformation on top of the previous ones (line 7) until the binary is marked as benign (line 5) or the maximum number of iterations is reached (line 2).

With this algorithm, a transformation is randomly selected at each iteration, and always applied. Hence, the baseline algorithm can require many iterations and oracle queries to turn the original malware into a misclassified binary.
Second, some transformations might suppress the effect of previous ones.
\rev{Third, the baseline algorithm considers each vendor equally good at detecting a malware, which is naive as the vendors display considerable diversity regarding detection strength. An algorithm that would target evasion on the strongest vendor first would increase the overall performance of the evasion process. }
Finally, we might generate a binary that entirely evades the oracle but is unpractical in terms of size or its execution performance. 

%Consequently, the transformation process should take into account the intrinsic attributes of the vendors. For instance, achieving total evasion may be expedited if stronger vendors are addressed earlier in the transformation process.

\begin{algorithm}

\SetKwData{optimizations}{optimizations}
\SetKwData{binary}{W}
\SetKwData{diversifier}{D}
\SetKwData{Malware Oracle}{MO}
\SetKwData{mutant}{E}
\SetKwInOut{Input}{input}\SetKwInOut{Output}{output}

\Input{binary $W$, diversifier $D$, Malware Oracle $MO$}
\Output{Benign binary $M'$ }
$M \leftarrow W$
\While{Not max iterations} {
    $M' \leftarrow D(M)$
    \eIf{ $MO(M')$ == "benign" } {
        \Return $M'$;
    } { 
        $M \leftarrow M'$;
    }
}
\Return "Not evaded";
\caption{Baseline evasion algorithm.}
\label{alg:naive}
\end{algorithm}

\subsubsection{MCMC Evasion}
% paragraph about the intuition the solution
To overcome the limitations of the baseline algorithm, we devise the \emph{MCMC evasion algorithm}, which we now discuss.
It is a Markov Chain Monte Carlo (MCMC) sampling  \cite{hastings1970monte} of the transformations to apply \cite{stoke}.
MCMC is used to sample from the space of transformation sequences to maximize the likelihood of oracle evasion in two ways.

The algorithm for the MCMC evasion is given in \autoref{alg:mcmc}.
The halting condition is met when a mutated binary is marked as benign, or a maximum number of iterations is reached.
The algorithm implements the MCMC in lines 6 to 15.
The Markov decision function in line 12 is used to determine whether it is worth applying the transformation at this step or whether it should be skipped.
This decision function is based on the current transformation's fitness and the fitness value saved at the previous step (line 14).
The core idea is to favor  a binary variant that evades the largest number of vendors  (lines 4 and 13).
Therefore, the number of oracle calls should decrease as the algorithm searches for transformations that converge toward total evasion.
On the other hand, if a new transformation step decreases the fitness value, it is likely to be ignored. 
The classical MCMC acceptance criteria in line 12 is meant to prevent the algorithm from being stuck in local minima.

% Sigma
In line 12, the fraction calculated in the exponentiation is controlled by the $\sigma$ parameter.
By setting a low $\sigma$ parameter, we can turn the MCMC evasion algorithm into a greedy algorithm.
In this case, the algorithm selects a new transformation only if the fitness value is higher than in the previous iteration.
On the contrary, if the $\sigma$ parameter is significant, the algorithm searches for local maxima.
In our experiments (\autoref{method}), we explain how we select the values of $\sigma$.

\rev{\ The MCMC evasion algorithm addresses the three limitations of the baseline mentioned in the previous section. First, the fitness function selects transformations, thus reducing the total number of transformations that are actually performed. Second, MCMC aims at increasing fitness, which reduces the risk of suppressing a valuable transformation performed in the previous step. Third, by favoring solutions that maximize the number of evaded vendors, MCMC biases the search towards evading the strongest vendors.}

\begin{algorithm}

\SetKwData{optimizations}{optimizations}
\SetKwData{binary}{W}
\SetKwData{diversifier}{D}
\SetKwData{mutant}{E}
\SetKwFunction{oracle}{O}
\SetKwInOut{Input}{input}\SetKwInOut{Output}{output}

\Input{binary $W$,  diversifier $D$, fitness function $FF$}
\Output{Benign binary $M'$ }

$M \leftarrow W$

previous\_fitness = $FF(W)$

\While{Not max iterations} {
    $M' \leftarrow D(M)$
    
    current\_fitness = $FF(M')$
        
    \eIf{ current\_fitness == 0 } {
        // Zero means that none vendor marks 
        
        // the binary as malware
        
        \Return $M'$;
    } { 
        $p \leftarrow random()$
        
        \If{ $p < min(1, exp(\sigma\frac{previous\_fitness}{current\_fitness}))$ } {
            $M \leftarrow M'$;
            
            previous\_fitness = current\_fitness;
        }

    }
    
}

\Return "Not evaded";

\caption{MCMC evasion algorithm.}
\label{alg:mcmc}
\end{algorithm}

\section{Experimental Methodology}
\label{method}

%\todo{R1 -  The RQ3 proposed to evaluate samples functionality, like an oracle, the same way detection was evaluated. However, there is no functionality oracle in the depicted figure from previous section.}

In this section, we enunciate the research questions around which we assess the ability of our technique to evade malware detectors. We also describe our dataset of malware, as well as the metrics we define to answer our research questions.

\subsection{Research Questions}

\begin{itemize}
    
    \item \textbf{RQ1. To what extent can \rev{cryptojacking} malware detection be bypassed by WebAssembly diversification?}
    With this research question, we evaluate the feasibility of binary transformations on malware and how they affect the detection of cryptojacking.

    \item \textbf{RQ2. To what extent can the attacker minimize the number of calls to the \rev{cryptojacking} detection oracle?} In real-world scenarios, the number of calls to the oracle is limited. With this question, we analyze the ability of our technique at limiting the number of oracles calls made during the evasion process.

    \item \textbf{RQ3. To what extent do the evasion techniques impact \rev{cryptojacking} malware functionality?} The evasion algorithms might generate variants that evade the detectors but modify their core malicious functionality. This research question evaluates the correctness of the created variants, as well as their efficiency.
    
    \item \textbf{RQ4. What are the most effective transformations for WebAssembly \rev{cryptojacking} malware evasion?} This research question provides empirical evidence on which types of transformations are better for WebAssembly cryptojacking evasion.

    \item \textbf{RQ5. To what extent can Wasm diversification evade \rev{cryptojacking} detection with MINOS?} In this research question, we evaluate the feasibility of our technique for evading  a state-of-the-art detector, MINOS, which is tailored to the analysis of WebAssembly.
    
\end{itemize}

\subsection{Dataset Selection}
\label{dataset}

To answer our research questions, we curate a dataset of WebAssembly malware.
For this, we filter the wasmbench dataset of Hilbig \etal \cite{Hilbig2021AnES} to collect suspicious malware according to VirusTotal. 
The wasmbench dataset contains 8643 binaries collected from GitHub repositories and web pages in 2021. 
To our knowledge, this dataset is the newest and most exhaustive collection of real-world WebAssembly binaries.
On August 2022, we passed the 8643 binaries of wasmbench to VirusTotal \cite{virustotal}, and we 33 binaries were marked as potentially dangerous by at least one antivirus vendor of VirusTotal.
\rev{All malware were marked as cryptojacking programs and we use these WebAssembly binaries to answer our research questions for cryptojakcing malware evasion.}

In \autoref{tab:the_33} we describe the 33 binaries detected as malware by VirusTotal. 
% Describes the table
%\todo{R1 - add the labels eg S in the text}
The table contains the following properties as columns: the identifier of the WebAssembly binary which is the sha256 hash of its bytestream, its size in bytes, the number of instructions, the number of functions defined inside the binary and the number of VirusTotal vendors that detect the binary. The last column contains the origin of the binary according to the wasmbench dataset.
\footnote{Binaries that could be found in a live webpage at the moment of this writing are marked with \live\ http:// .}

\footnotetext[2]{Yara project \url{https://github.com/davbo/yara-rs/tree/master/sample-miners}}
\footnotetext[3]{All Wasm binaries of the MinerRay project could be found at \url{https://github.com/miner-ray/miner-ray.github.io/tree/master/Data/SampleWasmFiles}}
\footnotetext[4]{All Wasm binaries of the SEISMIC project could be found at \url{https://github.com/wenhao1006/SEISMIC}}
\footnotetext[5]{Deepminer project \url{https://github.com/deepwn/deepMiner}}

% Add this in the reproduction subsection
%\todo{R1 - give a few more details about how you determine if the exectusion of the binary can be reproduced and how you indicate this in the last column} if the execution of the binary can be reproduced in an end-to-end environment. 

\begin{table}
    \caption{The 33 real-world WebAssembly cryptojacking used in our experiments. The table contains: the 256 hash of the WebAssembly cryptojacking, its size in bytes, the number of instructions, the number of functions defined inside the binary and the number of VirusTotal vendors that detect the binary at the time of writing. The last column contains the origin of the binary.}
    \label{tab:the_33}
    \footnotesize

\renewcommand\arraystretch{1.30}

\begin{tabularx}{\linewidth}{l|l ll  lp{0.40\linewidth}}
    \toprule
    
        \textbf{Hash}  & \textbf{S} & \textbf{\#I.} & \textbf{\#F.} & \textbf{\#D}  & \textbf{Origin} \\
\hline
% Result from VT
9d30e7f0  & 68796 & 30768 & 61 & 30  & http archive \\

% "Monero miner"
8ebf4e44  & 68803 & 30768 & 61 & 26 & Web crawling \\

47d29959 &  68796 & 30768 & 61 & 31 & Yara  \footnotemark[2] \\

% "Crytonight Miner"
 aafff587  
% cryptonight obfuscated add to the description of the group 
& 97551 & 47033 & 72 & 6 & SEISMIC \footnotemark[3]  \\

dc11d82d  & 67496 & 30246 & 49 & 20 & MinerRay \footnotemark[4] \\ 
% & & & & & \live \url{http://topfilmonline.ru} \\

% 64bits worker
0d996462 &  70972 & 30531 & 30 & 19  & SEISMIC, MinerRay \\ 
% & & & & & \live \url{http://filelar.com} \\

% 64bits worker instrumented

 fbdd1efa  & 94270 & 45905 & 40 & 18 &  SEISMIC \\

 a32a6f4b & 94461 & 45940 & 40 & 18 &  SEISMIC  \\

 d2141ff2  & 70111 & 31783 & 30 & 9 & MinerRay, SEISMIC \\
 
 046dc081  & 74099 & 31783 & 29 & 6 &  MinerRay, SEISMIC \\

24aae13a   & 62458 & 28339 & 37 & 4 &  SEISMIC \\

000415b2   
%    cryptonight obfuscated TODO check, this can be possibly merge with another group 
& 62466 & 28339 & 37 & 3  & SEISMIC \\

643116ff   & 73010 & 31866 &  & 6 & MinerRay  \\

    006b2fb6   & 87502 & 39544 & 90 & 4 & DeepMiner \footnotemark[5] \\

15b86a25   & 100755 & 45881 & 79 & 4 &  MinerRay \\

4cbdbbb1  
%NFWebMiner add to the description
% https://minero.cc/
& 104666 & 47916 & 62 & 3 & SEISMIC \\

 119c53eb  & 137320 & 67069 & 79 & 2   & SEISMIC \\

f0b24409 & 77572 & 34918 & 59 & 2 & MinerRay   \\

c1be4071 & 77572 & 34918 & 59 & 2 & MinerRay    \\

a74a7cb8 & 77572 & 34918 & 59 & 2 & MinerRay    \\

a27b45ef & 77572 & 34918 & 59 & 2 & MinerRay     \\

6b8c7899 & 77572 & 34918 & 59 & 2 & MinerRay   \\

68ca7c0e & 77572 & 34918 & 59 & 2 & MinerRay     \\

65debcbe & 77572 & 34918 & 59 & 2 & MinerRay    \\

5bc53343 & 77572 & 34918 & 59 & 2 & MinerRay    \\

59955b4c & 77572 & 34918 & 59 & 2 & MinerRay    \\

942be4f7    & 103520 & 46208 & 79 & 4 & MinerRay \\

fb15929f   & 77054 & 33562 & 112 & 4 &  MinerRay  \\

7c36f462   & 121931 & 55839 & 79 & 4 &  MinerRay \\

89a3645c   & 75003 & 34134 & 58 & 2 &  MinerRay \\

dceaf65b    
% cryptonight-(v1|v7)  Check that this is packaged with multiple versions
& 77575 & 34901 & 58 & 2 &   MinerRay  \\

089dd312   
% cryptonight-(v1|v7)  Check that this is packaged with multiple versions
& 79883 & 34989 & 58 & 2 & MinerRay \\ 
%& & & & &  \live \url{http://theroyalforums.com}\\
%& & & & &  \live \url{http://prylive.com.meta}  \\

e09c32c5  
%cryptonight (only \_cryptonight\_hash function) Check this 
& 71955 & 32416 & 46 & 1 & MinerRay  \\

\end{tabularx}
\end{table}

% General insights
The programs include between  30 and 70 functions, for a total number of instructions ranging from 30531 to 55839. 
The size of the programs ranges from 62 to 103 kilobytes.
These binaries are detected as malicious by at least 1 antivirus, and at most 31.
We have observed that  6 out of 33 binaries can be executed end-to-end.

To validate that the detected binaries are cryptojacking, we manually analyze  each of the 33 binaries identified as malign.
First, we observe that the binaries in the dataset originate from two primary sources: project SEISMIC and project MinerRay.
SEISMIC \cite{SEISMIC} is a research project about  instrumentation and monitoring at runtime to detect cryptojacking binaries.
MinerRay is also a research project to detect crypto mining processes in web browsers \cite{9286112}.
% It monitors the interaction between JavaScript and WebAssembly at runtime to detect crypto algorithm patterns.
Both projects have collected the binaries as real cryptojacking from the web and the dataset is a union of them.

Second, we observee that  all binaries share code from cryptonight  \cite{xmrig}, which is a library for cryptomining hashing. 
This observation is consistent with the findings of Romano \etal \cite{9286112}. 
We find 5 binaries that are multivariant packages \cite{MEWE} of cryptopnight. A multivariant package is a binary containing more than one hashing function.
Concretely, the binaries \texttt{0d996462}, \texttt{d2141ff2}, \texttt{046dc081}, \texttt{a32a6f4b} and \texttt{fbdd1efa} contain between 2 and 3 versions of hashing functions \texttt{cryptonight\_hash}.

%\todo{R1 - considered is vague here. are they different? did you compare them? static or dynamic? did Hilbig compare them? static or dynamic?}
Third, our manual analysis of the binaries reveals \thevariations\ main sources for these differences. 1) \textbf{Versions:} the binaries do not depend on the same version of cryptonight, 
2)  \textbf{Function reordering:} The order in which the functions are declared inside the binary changes 
3) \textbf{Innocuous expressions:} Expressions have been injected into the program code, but their execution does not affect the semantic of the original program 
% \footnote{The binaries offering this kind of variation can be found in the SEISMIC project. SEISMIC instruments 4 types of instructions to profile their execution. In addition, it includes 10 functions to interact with the profiling information. The massive injection of instructions for instrumenting explains how it almost doubles the size and the number of instruction of the first three binaries. These cases are annotated with \child\ symbol in \autoref{tab:the_33}}.
4) \textbf{Function renaming:} The name of the functions exported to JavaScript have been changed 
5) \textbf{Data layout changes:} The data needed by the cryptominers has different location in the WebAssembly linear memory. 
6) \textbf{Partial cryptonight:} Some binaries exclude cryptonight functions, i.e., the \texttt{cryptonight\_create}, \texttt{cryptonight\_destroy}, \texttt{cryptonight\_hash} functions.
It is interesting to note that changes in function order, function names, or data layout leads to different programs that have the same number of instructions and functions, as is the case for 9 of our malign binaries.

%\todo{R1 - Notice that MinerRay binaries have the same \#Instructions and \#Functions.}

\subsection{Methodology}

Based on our dataset of WebAssembly binaries, we follow the following procedures to answer our research questions.

\emph{RQ1: Evasion effectiveness.}
To answer RQ1, we execute the baseline evasion algorithm discussed in \autoref{sec:evasion-algorithms}. 
We first pass a suspicious binary to wasm-mutate.
The diversified version is passed to VirusTotal as a binary oracle.
If at least one vendor still detects the diversified binary, we pass it to wasm-mutate again to stack a new random transformation.
We repeat this process until VirusTotal does not label the binary as malware or reach a limit of 1000 transformations. 
The process is performed 10 times with 10 different random seeds for each binary.

\emph{RQ2: Oracle minimization.}
Malicious actors have a limited budget to perform evasion before they get caught.
The number of oracle calls is a proxy for such a budget, i.e., the lesser the number of oracle calls, the lesser effort spent.
To answer RQ2, we aim at minimizing the number of oracle calls while keeping the same evasion effectiveness.
In particular, we assess the capability of the MCMC evasion algorithm to minimize the number of calls to the malware oracle (\autoref{metric:oracle_calls}).
For this, we execute the MCMC evasion algorithm (\autoref{alg:mcmc}) one time for each malware binary of our dataset.
We use VirusTotal as an oracle and stop when we reach a limit of 1000 transformations. 
Since MCMC has a configuration parameter $\sigma$, we repeat the process  with three $\sigma$ values: $0.01$, $0.3$, and $1.1$.
The $\sigma$-value weights exploration and exploitation of transformations during the evasion process.
For example, the first value is low, favoring exploration at the most, meaning that the MCMC algorithm will take any new transformation, whether it increases the fitness function value or not. 
On the contrary, the largest value $1.1$ favors the exploitation and, meaning that during evasion, MCMC will only accept transformations with higher values from the oracle, i.e., more evaded detectors.
We manually select the third value $0.3$ as the balance between exploration and exploitation.

\emph{RQ3: Malware functionality.}
\label{rq3method}
% Motivation
A diversified binary that fully evades the detection, might not be practical due to behavioral or performance issues.
RQ3 complements our first two research questions with a correctness and an efficiency evaluation.
For every cryptojacking that can be executed, we reproduce all cryptominer components (described in \autoref{background:crypto}) and replace the WebAssembly binaries with variants  that fully evade VirusTotal.
\rev{For each executable cryptojacking program, we generate 10 variants with the baseline evasion algorithm as well as 10 variants with the MCMC algorithm, with $\sigma$ value $0.3$ to balance the MCMC exploration-exploitation. 
Then, we replace the original cryptojacking by each of the 20 variants in order to determine that the behavior of the original cryptojacking program is preserved in the variants \cite{demetrio2021functionality}.}

Our end-to-end pipelines provide data on the number of generated hashes in the webpage component as an HTML element.
Besides, the number of successful or incorrect jobs are logged by the miner pool.
This information can be used to measure both the correctness and efficiency of the generated WebAssembly variants. 
To collect data on the number of hashes per second, the webpage is accessed with Puppetter, while the miner pool logs are saved to measure the number of successful and failed jobs with their respective log time. 
By analyzing these two types of data, the overall correctness and effectiveness of a diversified WebAssembly binary can be determined.

%\todo{R1 - 
%- Still for RQ3, it is not clear how samples are instrumented to be validated. Is the code modified? Run in a debugger? In a sandbox? In other words, how the hash values are extracted from the the samples?
%}

% Correctness
To check correctness, we verify that the  hashes generated by the variants are valid.
We determine whether the hashes reach the third component of a cryptojacking (see \autoref{background:crypto}), i.e., how many successful and failed jobs are reported by the miner pool.
If a miner pool correctly accepts the  hashes, then the cryptojacking variant generated by our evasion algorithms is considered correct.

% efficiency
To check for efficiency, we measure the frequency of \texttt{hashes} produced by the variants binaries. 
For each WebAssembly cryptojacking and its generated variants, we execute and measure the hashes produced per second, during  1000 seconds.
For each malware variant, we check if the hash production frequency  is still in the same order of magnitude as the original.

\emph{RQ4: Individual Transformation Effectiveness.}
As discussed in \autoref{diversifier}, our diversifier comprises 135 possible transformation operators.
In this research question, we want to study which transformations perform better in evading the VirusTotal malware detection oracle.
This investigation will help future researchers and detectors engineers to improve their detection methods and tackle subversive transformations.

% How
We use a value of $\sigma=1.1$ to tune the MCMC evasion algorithm.
With this parameter, the MCMC evasion algorithm only keeps transformations that significantly contribute to improving the fitness of the variant.
In other words, a transformation is applied if at least one more detector of VirusTotal is evaded.
Then, we count the number of applied transformations, aggregated by its type.
By measuring this,  we obtain the most used one (resp. the least used), and, therefore, understand where malware researchers should focus for counter-evasion.

    \emph{RQ5: Effectiveness against MINOS} 
    This research question assesses the effectiveness of our evasion technique with respect to the state-of-the-art WebAssembly malware detector, MINOS \cite{minos}.
    This detector takes a Wasm binary as input and creates a 100x100 grayscale image from its pure byte stream.
    Using a Convolutional Neural Network, the generated image is classified as benign or malware.
    
    We replicate MINOS. 
    However, the model of MINOS is not publicly available, so we train our own model for this experiment.
    We use our own dataset to train the model with 33 malign programs and 33 benign programs.
    The 33 malign programs for training MINOS are the same listed in \autoref{tab:the_33}, the 33 benign programs are collected from the original MINOS reproduction steps.
    Our reproduction of MINOS achieves the same results as   the original paper, based on one-off validation. 
    Our reproduction of MINOS is publicly available at \url{https://github.com/ASSERT-KTH/ralph}.
%    \todo{TBD}
 %   We could not reproduce the full MINOS dataset, neither their way of collecting the binaries.
 %   Unfortunately, we tried to reach the authors several months ago asking for the dataset, and we have not received answers.

    We use MINOS as an oracle and follow the same method proposed in RQ1, passing each one of the binaries in \autoref{tab:the_33} to our diversifier and then to MINOS.
    For each binary, we do the process 10 times with 10 different seeds, generating a total of 330 variants with no more than 1000 stacked mutations.
    %\todo{TBD: Should we talk about that we could not fully reproduce (in terms of dataset)}

 %   \vspace{10mm}

%    \todo{let's discuss this last part}
 %    The answers to this question should also fill the gap in current arguments regarding the lack of obfuscators claimed by implemented Wasm detectors.

\subsection{Metrics}

In this section, we define the notions of total and partial evasion used in this work to measure the impact of the evasion algorithms proposed in \autoref{sec:evasion-algorithms}.
Besides, we also define the \texttt{number of oracle calls} and \texttt{number of stacked transformations} metrics. 
In addition, we define the metrics for correct hashes generated by the malware variants and the hashes generation speed.

% Notions of total and partial evasion
\begin{defi}{Total evasion:}\label{def:total_evasion}
Given a malware WebAssembly binary, an evasion algorithm generates a variant that totally evades detection if  \textit{all} detectors that originally identify the binary as malware identify the variant as benign.
\end{defi}

\begin{defi}{Partial evasion:}\label{def:partial}
Given a malware WebAssembly binary, an evasion algorithm generates a variant that partially evades detection if  \textit{at least one} detector that originally identifies the binary as malware identifies the variant as benign.
\end{defi}

% Oracle calss and stacked transformations
\begin{metric}{Number of oracle calls:}\label{metric:oracle_calls}
The number of calls made to the malware oracle during the evasion process.
\end{metric}

\begin{metric}{Number of stacked transformations:}\label{metric:stacked}
The total number of transformations applied on the initial malware binary during the evasion process.
\end{metric}

Notice that \autoref{metric:oracle_calls} is the number of times that lines 4 and  5 in \autoref{alg:naive} and \autoref{alg:mcmc} are executed, respectively. The same could be applied to \autoref{metric:stacked}, in lines 7 and 13 of \autoref{alg:naive} and \autoref{alg:mcmc}, respectively. 

The main purpose of a WebAssmbly cryptominer is to calculate hashes.
By measuring the number of calculated hashes per time unit, we can measure how performant the cryptojacking is.
Therefore, the impact of the evasion process over the performance of the created binary can be measured, calculating the number of hashes per time unit:

\begin{metric}{Crypto hashes per second (h/s):}\label{metric:hashes}
    Given a WebAssembly cryptojacking, the crypto hashes per second metric is the number of successfully generated hashes in one second.
\end{metric}

\begin{metric}{Correct crypto hashes:}\label{metric:correct}
    Given a WebAssembly cryptojacking, the number of correct crypto hashes is the number of hashes that the WebAssembly cryptojacking generates and that the miner pool accepts as valid.
\end{metric}

\section{Experimental Results}
\label{results}

In section, we answer our four research questions regarding the feasibility of WebAssembly diversification for malware evasion.

\subsection{RQ1. Evasion Effectiveness}

We run our baseline evasion algorithm with a limit of 1000 iterations per binary. 
At each iteration, we query VirusTotal to check if the new binary evades the detection.
This process is repeated with 10 random seeds per binary, resulting in a maximum of 10,000 queries per original binary.
In total, we generate 98714 variants for the original 33 suspicious binaries.

\autoref{tab:evasion_t1}  shows the data to answer RQ1.
The table contains as columns: the hash of the program, calculated as the sha256 hash, as its identifier, the number of initial VirusTotal detectors flagging the malware, the number of evaded antivirus vendors (cf. \autoref{def:partial}) and the mean number of iterations needed to generate a variant that fully evades the detection (cf. \autoref{def:total_evasion}).
The rows of the table are ordered with respect to the number of detectors for the original binary.

We observe that the baseline evasion algorithm successfully generates variants that totally evade detection for 30 out of 33 binaries.
The mean value of iterations needed to generate a variant that evades all detectors ranges from 120 to 635 stacked transformations.
For the 30 binaries that completely evade detection, we observe that the mean number of iterations to evade is correlated to the number of initial detectors.
For example, the \texttt{a32a6f4b} binary, initially flagged by 18  detectors, requires around 635 iterations, while the \texttt{309c32c5}, with only one initial flag, needs 120 iterations.
The mean number of iterations needed is always less than 1000 stacked transformations.

%\todo{R1 - Number of evaded detectors for those not completely evade.}
%\todo{R1 - 
%- In Table 2, what is the ID? I mean, is the program identifier a hash? which one?}

%\todo{R1 - - In terms of visualization, Table 2 would be better as a graph, so we could see better how detection and evasion rates are distributed.}
\begin{table}[t]

  \caption{Baseline evasion algorithm for VirusTotal. The table contains as columns: the hash of the program, the number of initial VirusTotal detectors, the maximum number of evaded antivirus vendors and the mean number of iterations needed to generate a variant that fully evades detection. The rows of the table are sorted by the number of initial detectors, from left to right and top to bottom. }
    \label{tab:evasion_t1}
  
\centering
\small
\renewcommand\arraystretch{1.10}
%\begin{adjustbox}{width=1\linewidth}
    \begin{tabular}{lrrrr}
        \toprule
         Hash &  \#D  &  Max. \#evaded  & Mean \#trans. \\
         %& & \textbf{hash} &  \textbf{\#D}  &  \textbf{Max. \#evaded} &   & \textbf{hash} &  \textbf{\#D}  &  \textbf{Max. \#evaded} & \\
        \hline
47d29959 &                 31 &             {\color{gray} 26} (83.8\%) &     N/A    \\ 
9d30e7f0 &                 30 &             {\color{gray} 24} (80.0\%) &      N/A    \\ 
8ebf4e44 &                 26 &             {\color{gray} 21} (80.7\%)&     N/A     \\

\hline 
dc11d82d &                 20 &       20 (100.0\%)       &  355  \\ 
0d996462 &                 19 &     19 (100.0\%)     &  401  \\ 
a32a6f4b &                 18 &       18 (100.0\%)       &  635 \\

fbdd1efa &                 18 &         18 (100.0\%)     &  310 \\ 
d2141ff2 &                  9 &          9 (100.0\%)     &  461  \\ 
aafff587 &                  6 &          6 (100.0\%)     &  484  \\

046dc081 &                  6 &          6 (100.0\%)     &  404 \\ 
643116ff &                  6 &          6 (100.0\%)     &  144  \\ 
15b86a25 &                  4 &          4 (100.0\%)     &  253 \\

006b2fb6 &                  4 &           4 (100.0\%)    &  282 \\ 
942be4f7 &                  4 &           4 (100.0\%)    &  200 \\ 
7c36f462 &                  4 &           4 (100.0\%)    &  236 \\

fb15929f &                  4 &            4 (100.0\%)   &  297 \\ 
24aae13a &                  4 &         4 (100.0\%)      &  252 \\ 
000415b2 &                  3 &         3 (100.0\%)      &  302 \\

4cbdbbb1 &                  3 &          3 (100.0\%)     &  295 \\ 
65debcbe &                  2 &          2 (100.0\%)     &  131  \\ 
59955b4c &                  2 &          2 (100.0\%)     &  130 \\

89a3645c &                  2 &           2 (100.0\%)    &  431 \\
a74a7cb8 &                  2 &           2 (100.0\%)    &  124 \\
119c53eb &                  2 &           2 (100.0\%)    &  104 \\

089dd312 &                  2 &           2 (100.0\%)    &  153 \\
c1be4071 &                  2 &           2 (100.0\%)    &  130 \\
dceaf65b &                  2 &           2 (100.0\%)    &  140 \\

6b8c7899 &                  2 &            2 (100.0\%)   &  143 \\
a27b45ef &                  2 &         2 (100.0\%)      &  145 \\
68ca7c0e &                  2 &         2 (100.0\%)      &  137  \\

f0b24409 &                  2 &         2 (100.0\%)      &  127  \\
5bc53343 &                  2 &         2 (100.0\%)      &  118  \\
e09c32c5 &                  1 &         1 (100.0\%)      &  120 \\
    \end{tabular}
%\end{adjustbox}
\end{table}

%\todo{ An important information missing for RQ1 is how long it takes to generate the baseline variants. The authors even mention that " However, having more iterations seems not a realistic scenario", but do not tell the reader how long is this impractical.}

%\todo{R1 - - Authors could also discuss if the impracticability comes from the number of queries per itself, or due to the delay imposed by the VT queries. In other words, if the oracle was local and not remote, would the threshold be different?}

\begin{figure}
  \includegraphics[width=1.0\linewidth]{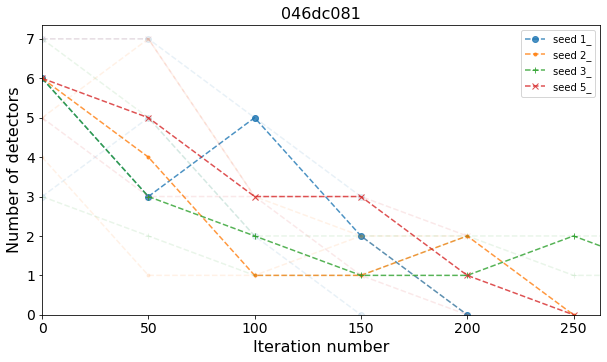}
  \caption{The figure shows the evasion process with four seeds for binary \texttt{046dc081}.
  Each point in the x-axis represents 50 iterations, and the y-axis represents the number of VirusTotal detectors.}
  \label{plot:progress}
\end{figure}

\autoref{plot:progress}  shows the evasion process with four different seeds for the binary \texttt{046dc081}.
Each point in the x-axis represents 50 iterations, and the y-axis represents the number of VirusTotal detectors flagging the binary. 
Three out of 4 seeds manage to totally evade VirusTotal in less than 250 iterations.
We have observed that there are better evasion techniques than pure random transformations.
For example, the seed represented by the green line partially evades the oracle but shows no tendency to evade detection before 300 iterations.
Besides, some transformations help some classifiers to detect the mutated binary.
These phenomena are empirically exemplified in \autoref{plot:progress} in which the curves is not always monotonously decreasing, like the blue-colored curve. 
In this case, it goes from 3 VirusTotal detectors to 5 during the 50-100 iterations.

% The cases in which we could not, partial evasion
There are 3 binaries for which the baseline algorithm does not completely evade the detection. 
In these three cases, the algorithm misses 5 out 31, 6 out of 30 and 5 out 26 detectors.
The  explanation is the maximum number of iterations (1000) we use for our experiments.
However, having more iterations seems not a realistic scenario.
For example, if some transformations increment the binary size during the transformation, a considerably large binary might be impractical for bandwidth reasons.

On the other hand, there is a balance between generating variants and avoiding detection by defense mechanisms.
For example, VirusTotal detects when it is being stressed with too many requests and too many queries can be detected and blocked, effectively ruining evasion. 
In our attack scenario, where VirusTotal is used as an external detector (see \autoref{attack-model-section}), we must be mindful of this limit.
In contrast, when defense mechanisms such as MINOS are placed locally, we believe that the number of queries is not necessarily limited. 

%\todo{R1 - Authors should also clarify if the transformations are idempotent or not. In other words, can I apply the transformations in any other or not?}
%\todo{R1 - In case I can apply transformation in any other, would paralelization help in increasing the number of transformations in a practical time slot?}

% Idemponent or not

Wasm-mutate performs mutations based on the input binary. 
In the experiments for RQ1, the input binaries for the baseline algorithm comes from the application of a previous mutation. 
Yet, we have observed that some transformations can be applied in any order. 
This means that different sequences of transformations can produce the same binary variant. 
This often happens when two mutation targets inside the binary are different, such as two disjoint pieces of code.
Therefore, a potential parallelization for the baseline algorithm is possible as soon as transformation sequences do not interfere with others.

Overall, our experiments prove that wasm-mutate is a powerful tool to perform malware evasion. 
By carefully selecting the order and type of transformations applied, it is possible to generate program variants that are both performant and effective at evading detection.
This same idea is explored in the next section.

\begin{tcolorbox}[boxrule=1pt,arc=.3em,boxsep=-1.3mm]
  \textbf{Answer to RQ1}: The baseline evasion algorithm with wasm-mutate clearly decrease the detection rate by VirusTotal antivirus vendors for cryptojacking malware.
  We achieve total evasion of WebAssembly cryptojacking malware in 30/33 (90\%) of our malware dataset.

\end{tcolorbox}

\subsection{RQ2. Oracle Minimization}

%\todo{R1 -  Still in RQ2, What are the qualitative results authors get from this experiment? In other words, is there a combination of transformation that is effective for all (or most) samples? If yes, why these transformations are of particular impact on AVs?}
%\todo{TBD: I think this should be discussed in RQ4}
% Recall the protocol
With RQ2, we analyze the effect of the MCMC evasion algorithm in minimizing the number of calls to the malware oracle (\autoref{metric:oracle_calls}).
To answer RQ2, we execute the MCMC evasion algorithm discussed in \autoref{alg:mcmc}.

In \autoref{tab:evasion_t2} we can observe the impact of the MCMC evasion algorithm on our reference dataset.
The first two columns of the table are the original program's hash and the number of initial VirusTotal detectors flagging the malware. 
The remaining columns are divided into two categories, maximum detectors evaded (\autoref{def:partial}) and the number of oracle calls if total (\autoref{def:total_evasion}).
Each one of the two categories contains the result for both evasion algorithms, first the baseline algorithm(BL) followed by the three $\sigma$-values analyzed in the MCMC evasion algorithm.
Notice that, for the baseline algorithm, the number of oracle calls is the same value as the number of transformations needed to evade by construction. 
We highlight in bold text the values for which the baseline or the MCMC evasion algorithms are better than each other, the lower, the better.

% General stats
We observe that the MCMC evasion algorithm successfully generates variants that totally evade the detection for 30 out of 33 binaries, it thus as good as the baseline algorithm.
The improvement happens in the number of oracle calls.
The oracle calls needed for the MCMC evasion algorithm are 92\% of the needed on average for the baseline evasion algorithm.

\begin{table}[t]

  \caption{MCMC evasion algorithm for VirusTotal. The first two columns of the table are: the identifier of the original program and the number of initial detectors. 
The remaining columns are divided into two categories, maximum detectors evaded if partial evasion  and number of oracle calls if total evasion.
Each one of the two categories contains the result of the evasion algorithms, first the baseline algorithm (BL) followed by the three $\sigma$-values analysed in the MCMC evasion algorithm.
We highlight in bold text the values for which the baseline or the MCMC evasion algorithms are better from each other. Overall, the MCMC evasion algorithm needs less oracle calls than the baseline algorithm.}
    \label{tab:evasion_t2}
  
\large

\renewcommand\arraystretch{1.20}
\begin{adjustbox}{width=1\linewidth}
  \begin{tabular}{p{1.55cm}|r|r|ccc|r|ccc}
          \toprule

Hash &  \#D & \multicolumn{4}{c|}{Max. evaded} & \multicolumn{4}{c}{\#Oracle calls} \\
\hline
 &  & BL &  \multicolumn{3}{c|}{MCMC}  & BL & \multicolumn{3}{c}{MCMC} \\
 \hline
 & & & $\sigma$=0.1 & $\sigma$=0.3 & $\sigma$=1.1 & & $\sigma$=0.01 & $\sigma$=0.3 & $\sigma$=1.1 \\
\hline
47d29959 & 31 & \textbf{26} & 19 & 12 & 10 &  &  &  &  \\
9d30e7f0 & 30 & \textbf{24} & 17 & 9 & 10 &  &  &  &  \\
8ebf4e44 & 26 & \textbf{21} & 13 & 5 & 4 &  &  &  &  \\
\hline
dc11d82d & 20 & \textbf{20} & \textbf{20} & 14 & 15 & \textbf{355} & 446 &  &  \\
0d996462 & 19 & \textbf{19} & \textbf{19} & 14 & 4 & \textbf{401} & 697  &  &  \\
a32a6f4b & 18 & \textbf{18} & \textbf{18} & 6 & 1 & 635 & \textbf{625}  &  &  \\
fbdd1efa & 18 & \textbf{18} & \textbf{18} & 3 & 3 & \textbf{310} & 726 &  &  \\
d2141ff2 & 9 & \textbf{9} & \textbf{9} & \textbf{9} & 5 & \textbf{461} & 781  & 783  &  \\
aafff587 & 6 & \textbf{6} & 2 & \textbf{6} & \textbf{6} & 484 &  & \textbf{331} & 413 \\
046dc081 & 6 & \textbf{6} & \textbf{6} & \textbf{6} & 5 & 404 & 397 & \textbf{159} &  \\
643116ff & 6 & \textbf{6} & \textbf{6} & \textbf{6} & 1 & \textbf{144} & 436  & 631  &  \\
15b86a25 & 4 & \textbf{4} & \textbf{4} & \textbf{4} & \textbf{4} & 253 & 208 & 214 & \textbf{131} \\
006b2fb6 & 4 & \textbf{4} & \textbf{4} & \textbf{4} & 0 & \textbf{282} & 380  & 709  &  \\
942be4f7 & 4 & \textbf{4} & \textbf{4} & \textbf{4} & \textbf{4} & \textbf{200} & \textbf{200} & \textbf{200} & 219 \\
7c36f462 & 4 & \textbf{4} & \textbf{4} & 2 & 0 & 236 & \textbf{221} &  &  \\
fb15929f & 4 & \textbf{4} & 2 & \textbf{4} & 2 & \textbf{297} &  & 475  &  \\
24aae13a & 4 & \textbf{4} & \textbf{4} & \textbf{4} & 0 & \textbf{252} & 401  & 446 &  \\
000415b2 & 3 & \textbf{3} & \textbf{3} & \textbf{3} & 2 & 302 & 376 & \textbf{34} &  \\
4cbdbbb1 & 3 & \textbf{3} & \textbf{3} & \textbf{3} & \textbf{3} & 295 & 204 & \textbf{72} & 685 \\
65debcbe & 2 & \textbf{2} & \textbf{2} & \textbf{2} & \textbf{2} & 131 & \textbf{33} & \textbf{33} & \textbf{33} \\
59955b4c & 2 & \textbf{2} & \textbf{2} & \textbf{2} & \textbf{2} & 130 & \textbf{33} & \textbf{33} & \textbf{33} \\
89a3645c & 2 & \textbf{2} & \textbf{2} & \textbf{2} & \textbf{2} & 431 & 319  & 197 & \textbf{107} \\
a74a7cb8 & 2 & \textbf{2} & \textbf{2} & \textbf{2} & \textbf{2} & 124 & \textbf{33} & \textbf{33} & \textbf{33} \\
119c53eb & 2 & \textbf{2} & \textbf{2} & \textbf{2} & \textbf{2} & 104 & 45  & 480  & \textbf{18}  \\
089dd312 & 2 & \textbf{2} & \textbf{2} & \textbf{2} & \textbf{2} & 153 & 166  & 167 & \textbf{123} \\
c1be4071 & 2 & \textbf{2} & \textbf{2} & \textbf{2} & \textbf{2} & 130 & \textbf{33} & \textbf{33} & \textbf{33} \\
dceaf65b & 2 & \textbf{2} & \textbf{2} & \textbf{2} & \textbf{2} & 140 & 166  & 166  & \textbf{132} \\
6b8c7899 & 2 & \textbf{2} & \textbf{2} & \textbf{2} & \textbf{2} & 143 & \textbf{33} & \textbf{33} & \textbf{33} \\
a27b45ef & 2 & \textbf{2} & \textbf{2} & \textbf{2} & \textbf{2} & 145 & \textbf{33} & \textbf{33} & \textbf{33} \\
68ca7c0e & 2 & \textbf{2} & \textbf{2} & \textbf{2} & \textbf{2} & 137 & \textbf{33} & 167 & 595 \\
f0b24409 & 2 & \textbf{2} & \textbf{2} & \textbf{2} & \textbf{2} & 127 & \textbf{11} & 33 & \textbf{11} \\
5bc53343 & 2 & \textbf{2} & \textbf{2} & \textbf{2} & \textbf{2} & 118 & \textbf{33} & \textbf{33} & \textbf{33} \\
e09c32c5 & 1 & \textbf{1} & \textbf{1} & \textbf{1} & 0 & \textbf{120} & 488  & 921  
&  \\
\end{tabular}

\end{adjustbox}
\end{table}

% The best cases
For 21 of 30 binaries that evade detection entirely, we observe that the mean number of oracle calls needed is lower than those in the baseline evasion algorithm.
For example, \texttt{f0b24409} needs 11 oracle calls with the MCMC evasion algorithm to fully evade VirusTotal, while for the baseline evasion algorithm, it needs 127 oracles calls.
For those 21 binaries, it needs only 40\% of the calls the baseline evasion algorithm needs.

\begin{figure}
  \includegraphics[width=1.0\linewidth]{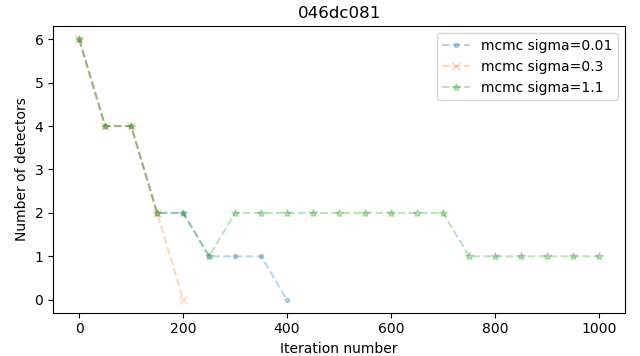}
  \caption{The figure shows the MCMC evasion process for the binary \texttt{046dc081}.
  Each point in the x-axis represents 50 iterations, and the y-axis represents the number of VirusTotal detectors. The figure shows less chaotic progress for oracle evasion. }
  \label{plot:comp}
\end{figure}

The impact of the MCMC evasion algorithm is illustrated in \autoref{plot:comp}.
Each point in the x-axis represents 50 iterations, and the y-axis represents the number of VirusTotal detectors flagging binary 046dc081.
2 out of 3 $\sigma$-values manage to totally evade VirusTotal in less than 400 iterations.
On the contrary, lower acceptance criteria $\sigma=1.1$ (green line) partially evades the oracle, but does not fully evade within the maximum 1000 iterations limit of the experiment.

% Sigma values
The $\sigma$ value in the \autoref{alg:mcmc} provides the acceptance criteria for new transformations in the MCMC evasion algorithm.
We run the MCMC evasion algorithm with three values to better understand the extent to which the transformation space is explored to find a binary that evades successfully.
We have observed that for a large value of $\sigma$, meaning low acceptance, 16 binaries cannot be mutated to evade VirusTotal entirely.
The main reason is that, in this case, the MCMC evasion algorithm discards transformations that increase the number of oracle calls.
Therefore, the algorithm gets stuck in local minima and never finds a binary that entirely evades.
On the contrary, a small value of $\sigma$ accepts more new transformations even if more detectors flag the binary.

% There is no best sigma
According to our experiments, $\sigma=0.3$ offer a good acceptance trade-off.
However, the best value of $ \sigma $ actually depends on the binary to mutate.
Therefore, we cannot conclude the best value of $\sigma$ for the whole dataset. 
This is a consequence of the particularities of each one of the original binaries and its detectors.
For example, we have observed that for the bottom part of \autoref{tab:evasion_t2} the highest value of $\sigma$ works better overall.
The main reason is the low number of original detectors. 
In those cases, the exploration space for transformations is smaller.
Thus, for the MCMC evasion algorithm, the chances of a local minimum for the fitness function to be global are larger.
Therefore, the MCMC evasion algorithm with low acceptance criteria can find the binary that fully evades in fewer iterations.
On the contrary, if the number of initial detectors is more significant, the exploration space is too big to explore in 1000 max iterations.
The reason is that the MCMC evasion algorithm applies one transformation per time. 
While we provide fine-grained analysis in our work,
more than one transformation per iteration could be applied in the MCMC evasion algorithm to solve this.

% Max evaded
In all 3 cases for which neither the baseline nor the MCMC evasion algorithms could find a binary that fully evades, the maximum evaded detectors are less for the MCMC evasion algorithm.
The main reason is that the MCMC evasion algorithm might prevent transformations for which the number of detectors increases.
As previously discussed, it is stuck in local minima, which means that it does not explore transformation paths for which a higher number of detectors could lead to better long-term results and, eventually, full evasion.

\begin{tcolorbox}[boxrule=1pt,arc=.3em,boxsep=-1.3mm]
  \textbf{Answer to RQ2}: 
The MCMC evasion algorithm needs
fewer oracle calls than the baseline algorithm. In 21 cases out of 33, it needs only 40\% of oracle calls compared to the baseline, providing more stealthiness to the malicious organization directing the evasion.
  The acceptance criterion $\sigma$ of the MCMC evasion algorithm needs to be carefully crafted depending on the original binary.
\end{tcolorbox}

\subsection{RQ3. Malware Functionality}
%\todo{R1 -  In RQ3, authors state that: "we select the six binaries we can build
%and execute end-to-end". What do end-to-end mean? Does it means that authors have access to the three components for these samples? Did authors set a web server to perform the infection in this case? 
%This is the first time authors mention in the paper that they could not evaluate the correctness of all samples. This is a significant limitation.
%}

To answer RQ3, we select the six binaries we can build and execute end-to-end, because we have access to the three components previously mentioned in \autoref{background:crypto}.
For those six binaries, we are able to replace the original WebAssembly binary with variants generated by our evasion algorithms. 
% Although the webpages used in our reproduction are purely demonstrative, it is important to note that the injection of cryptojacking malware is feasible in real-world scenarios \cite{9566204}. For example, by using malvertising mechanism.

We execute the original binary and the variants generated by the baseline and MCMC evasion algorithms. The essence of a cryptojacking is to generate hashes at high-speed. Consequently, we assess the correctness of the variants concerning two properties: validity of the hashes and frequency of hash generation. With \autoref{metric:correct}, we determine the validity of the generated hashes by checking if the backend miner pool accepts them.
The frequency is measured as the amount of \texttt{hashes} produced by the variant binaries in one second (\autoref{metric:hashes}).

\autoref{tab:rq3} summarizes the key data for RQ3.
Each row of the table corresponds to one binary that can be executed end-to-end, by reproducing the three components mentioned in \autoref{background:crypto}.
The first two values of each row are the original binary's identifier and its original frequency of hash generation.
Then, for our baseline algorithm, each row has the data for \rev{10} variants generated during a successful oracle evasion.
For each one of the variants, we include the correctness percentage and the hashes calculated per second.
The last group of columns of the table contains the correctness percentage and the hashes calculated per second for the \rev{10} variants generated with the MCMC algorithm.
After each frequency of hashing measurement, we indicate the relative difference with the original frequency in parentheses.
A difference larger or equal to 1.0 means that the variant is faster or as efficient as the original. On the contrary, the variant is slower if the difference is lower than 1.0.
The table contains the information for 120 variants.

%\todo{R1 -  It is not clear why performance is increased with the transformation and not degraded. This is not what is observed in similar works. Performance is expected to degrade because as more instructions are execution, the more CPU cycles it takes. Please, explain.}

\begin{table*}[t]

      \caption{\rev{Malware correctness and efficiency.  
    We execute each cryptojacking that can be reproduced, with the original malware, and with 10 baseline variants and 10 MCMC variants.
    The first left section indicates the identifier of the original binary and its original frequency of hash generation.
    The second section of the table shows  correctness percentage and hashes per second for ten variants generated with our baseline algorithm.
    The third section of the table shows  correctness percentage and hashes per second for ten variants generated with the MCMC algorithm.
    After each frequency of hashing measurement, we indicate the relative difference with the original frequency, in parentheses.
    A difference larger or equal to 1.0 indicates a variant that is faster or as efficient as the original; a difference lower than 1.0 is a variant slower  than the original.}
     }
    \label{tab:rq3}

\renewcommand\arraystretch{1.2}
\begin{adjustbox}{width=1\linewidth}
\rev{
  \begin{tabular}{c c | cr  cr  cr || cr  cr  cr}

        \toprule
           Hash &  Original h/s & \multicolumn{6}{c||}{Baseline algorithm} & \multicolumn{6}{c}{MCMC algorithm} \\
         \hline

        0d996462  &  116.0     &  100\%  & 25 (0.22)   &  100\%  & 24 (0.21)   &  100\%  & 26 (0.22)    &  100\%  & 116 \textbf{(1.00)}    &  100\%  & 70 (0.60)   &  100\%  & 67 (0.58)  \\ &    &  100\%  & 116 \textbf{(1.00)}    &  100\%  & 110 (0.95)   &  100\%  & 30 (0.26)    &  100\%  & 110 (0.95)   &  100\%  & 76 (0.66)   &  100\%  & 60 (0.52)  \\ &    &  100\%  & 55 (0.47)   &  100\%  & 27 (0.23)   &  100\%  & 23 (0.20)    &  100\%  & 86 (0.74)   &  100\%  & 60 (0.52)   &  100\%  & 72 (0.62)  \\ &    &  100\%  & 27 (0.23)    &   &   &   &   &   100\%  & 76 (0.66)  \\
        \hline
 a32a6f4b  &  48.0   &   100\%  & 25 (0.52)   &  100\%  & 24 (0.50)   &  100\%  & 24 (0.50)    &  100\%  & 26 (0.54)   &  100\%  & 45 (0.94)   &  100\%  & 41 (0.85)  \\ &    &  100\%  & 26 (0.54)   &  100\%  & 25 (0.52)   &  100\%  & 26 (0.54)    &  100\%  & 46 (0.96)   &  100\%  & 41 (0.85)   &  100\%  & 45 (0.94)  \\ &    &  100\%  & 26 (0.54)   &  100\%  & 24 (0.50)   &  100\%  & 25 (0.52)    &  100\%  & 44 (0.92)   &  100\%  & 42 (0.88)   &  100\%  & 45 (0.94)  \\ &    &  100\%  & 23 (0.48)    &   &  &  &  &  100\%  & 45 (0.94)  \\
        \hline
 fbdd1efa  &  37.0     &  100\%  & 25 (0.68)   &  100\%  & 25 (0.68)   &  100\%  & 25 (0.68)    &  100\%  & 28 (0.76)   &  100\%  & 47 \textbf{(1.27)}    &  100\%  & 48 \textbf{(1.30)}   \\ &    &  100\%  & 25 (0.68)   &  100\%  & 26 (0.70)   &  100\%  & 26 (0.70)    &  100\%  & 47 \textbf{(1.27)}    &  100\%  & 47 \textbf{(1.27)}    &  100\%  & 53 \textbf{(1.43)}   \\ &    &  100\%  & 25 (0.68)   &  100\%  & 25 (0.68)   &  100\%  & 25 (0.68)    &  100\%  & 48 \textbf{(1.30)}    &  100\%  & 48 \textbf{(1.30)}    &  100\%  & 49 \textbf{(1.32)}   \\ &    &  100\%  & 25 (0.68)     &   &  &  &   &  100\%  & 47 \textbf{(1.27)}   \\
        \hline
   d2141ff2  &  113.0    &  100\%  & 54 (0.48)   &  100\%  & 55 (0.49)   &  100\%  & 55 (0.49)    &  100\%  & 107 (0.95)   &  100\%  & 107 (0.95)   &  100\%  & 107 (0.95)  \\ &    &  100\%  & 57 (0.50)   &  100\%  & 56 (0.50)   &  100\%  & 56 (0.50)    &  100\%  & 109 (0.96)   &  100\%  & 106 (0.94)   &  100\%  & 100 (0.88)  \\ &    &  100\%  & 57 (0.50)   &  100\%  & 53 (0.47)   &  100\%  & 53 (0.47)    &  100\%  & 101 (0.89)   &  100\%  & 100 (0.88)   &  100\%  & 107 (0.95)  \\ &    &  100\%  & 55 (0.49)     &   &  &  &   &  100\%  & 107 (0.95)  \\
        \hline
   046dc081  &  118.0    &  100\%  & 58 (0.49)   &  100\%  & 60 (0.51)   &  100\%  & 59 (0.50)    &  100\%  & 118 \textbf{(1.00)}    &  100\%  & 120 \textbf{(1.02)}    &  100\%  & 119 \textbf{(1.01)}   \\ &    &  100\%  & 60 (0.51)   &  100\%  & 55 (0.47)   &  100\%  & 62 (0.53)    &  100\%  & 120 \textbf{(1.02)}    &  100\%  & 116 (0.98)   &  100\%  & 120 \textbf{(1.02)}   \\ &    &  100\%  & 55 (0.47)   &  100\%  & 50 (0.42)   &  100\%  & 57 (0.48)    &  100\%  & 119 \textbf{(1.01)}    &  100\%  & 120 \textbf{(1.02)}    &  100\%  & 119 \textbf{(1.01)}   \\ &    &  100\%  & 55 (0.47)     &   &  &  &   &  100\%  & 120 \textbf{(1.02)}   \\
        \hline
   006b2fb6  &  8.0    &  100\%  & 7 (0.88)   &  100\%  & 6 (0.75)   &  100\%  & 4 (0.50)    &  100\%  & 6 (0.75)   &  100\%  & 6 (0.75)   &  100\%  & 6 (0.75)  \\ &    &  100\%  & 9 \textbf{(1.12)}    &  100\%  & 6 (0.75)   &  100\%  & 4 (0.50)    &  100\%  & 6 (0.75)   &  100\%  & 6 (0.75)   &  100\%  & 6 (0.75)  \\ &    &  100\%  & 4 (0.50)   &  100\%  & 6 (0.75)   &  100\%  & 4 (0.50)    &  100\%  & 8 \textbf{(1.00)}    &  100\%  & 9 \textbf{(1.12)}    &  100\%  & 6 (0.75)  \\ &    &  100\%  & 6 (0.75)     &   &  &  &  &  100\%  & 6 (0.75)  \\
        \hline

    \end{tabular}
}
\end{adjustbox}

\end{table*}

% All behave identically
We use the backend miner pools (see \autoref{background:crypto}) of the six cryptojacking to determine the validity of the hashes computed by all the programs.
All correctness assessments for the \rev{120} variants indicate that the miner pools do not detect any invalid hash when executing the WebAssembly cryptominer variants.
This means that the variants synthesized by the baseline and the MCMC algorithms to evade the malware oracle can still systematically generate valid hashes.

% Efficiency
We have observed that \rev{23 of 120} variants are more efficient than the original cryptojacking.
For example, for the \texttt{fbdd1efa} binary, all its variants are from 1.27 to 1.43 faster than the original hash generation frequency.
This phenomenon occurs because wasm-mutate can perform transformations in the executable code, which work as optimizations.
Our experiments have revealed two sources of faster WebAssembly variants: loop unrolling transformations and code replacements that lead to smaller binaries. 
We have also found that debloating transformations, which remove unneeded structures and dead code, results in more hashes being produced by the cryptominer in the first few seconds of mining, likely because of faster compilation.

In summary, all this is evidence that focused optimization is a good primitive for evasion. 
% The contrary case.
On the contrary, \rev{97 out of 120} variants underperform compared to the original binary.
The worst case is the binary \texttt{0d996462}.
Its slowest variant has \rev{0.20} of the original generation frequency.
The main reason is that wasm-mutate also introduces non-optimal transformations regarding performance.

The variants generated by the baseline evasion algorithm tend to be slower than the MCMC evasion algorithm.
The MCMC evasion algorithm triggers fewer oracle calls and can generate variants faster than the ones generated by the evasion algorithm.
This phenomenon is a direct consequence of the MCMC evasion algorithm implementation, which has a selective strategy when applying transformations on a binary  (see \autoref{alg:mcmc}).
In summary, the MCMC evasion algorithm produces variants that fully evade the VirusTotal oracle with lower performance overhead during execution. The worst performant variant is only 1.93 times slower for the MCMC evasion algorithm.

\begin{figure}
    \centering
  \includegraphics[width=0.7\linewidth]{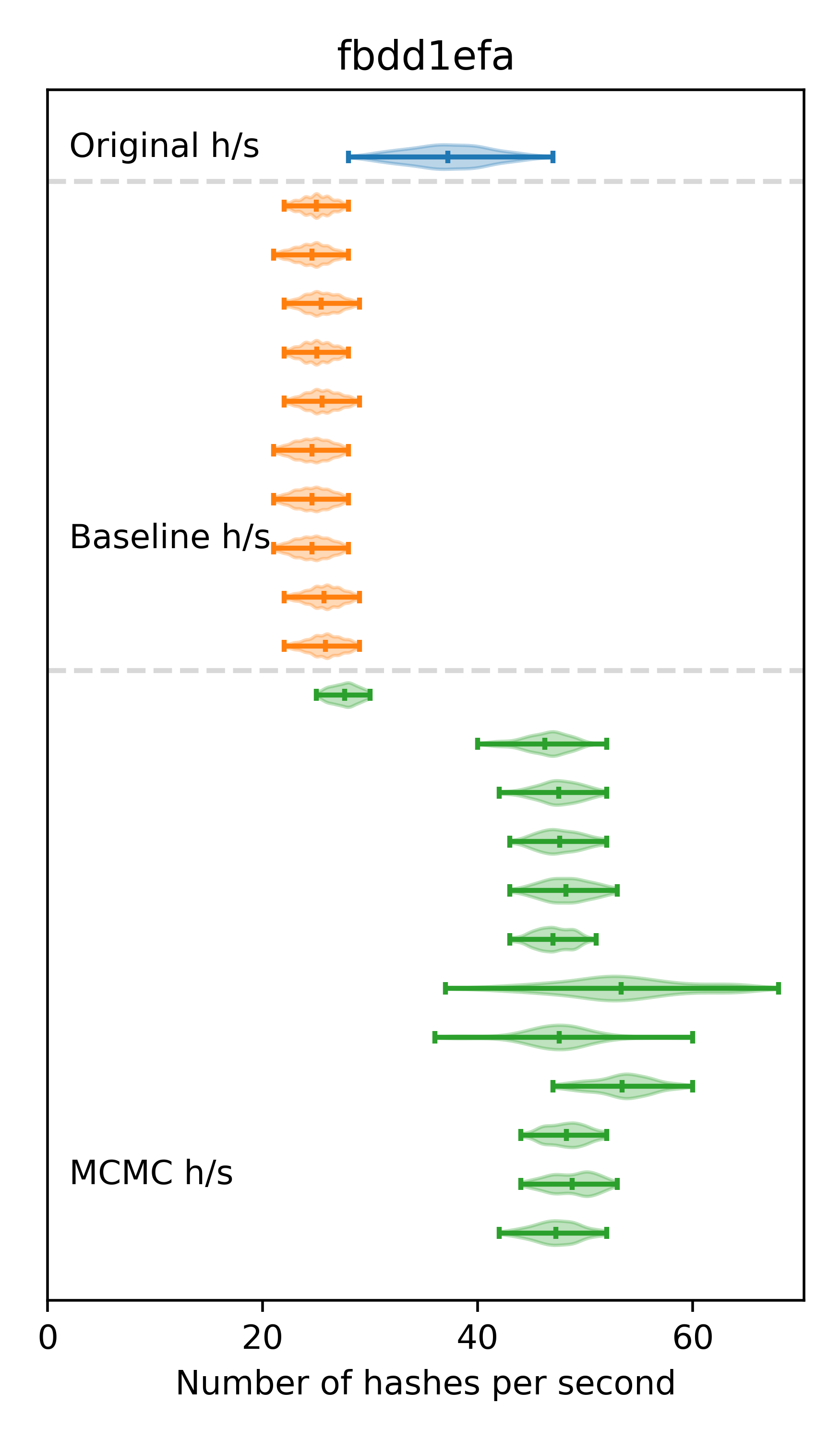}
  \caption{Distribution of the number of hashes per second for the variants generated for the original \texttt{fbdd1efa} cryptojacking.
In the figure we include the original binary (in blue), ten variants generated by the baseline algorithm (in orange), and ten variants generated by the MCMC evasion algorithm (in green).}
  \label{plot:distrib}
\end{figure}

% Instantiate with one binary
In \autoref{plot:distrib}, we plot the distribution of the hashes per second for variants generated for the  \texttt{fbdd1efa} cryptojacking.
In the figure, we include the original binary (in blue), the ten variants generated by the baseline algorithm (in orange), and the ten variants generated by the MCMC evasion algorithm (in green). 
Each violin plot corresponds to the hashes per second.
We observe a normal distribution around the exact number of hashes per second. 
In this case, we have observed that the MCMC evasion algorithm provides a hash-per-second ratio better than the original.
This phenomenon can be observed in the lines inside the violin plots, the green line is shifted to the right compared to the lines of the blue violin plot.

\begin{tcolorbox}[boxrule=1pt,arc=.3em,boxsep=-1.3mm]
  \textbf{Answer to RQ3}: 
   Our algorithms synthesize WebAssembly cryptojacking variants that fully evade our malware oracle and that provide the same functionality as the original. The execution of evading malware systematically produces valid hashes, and the variations in performance are imperceptible. For 19\% of the generated
variants, we observe better performance, and in the
worst case, the generated variant underperforms by five
times the original binary.
\end{tcolorbox}

\subsection{RQ4. Individual Transformation Effectiveness}

%\todo{R1 -  In 5.4, authors state that: "On the other hand, defenders can focus on the most effective transformations to evade antiviruses.". This is a vague statement. How would defenders use this information? I suppose adversarial retraining would be a possible answer for that, but it is not clarified. In this sense, extend the experiments to cover adversarial retraining would be a significant way to increase the contribution presented in this work.}

With RQ4, we investigate what individual transformations are the most appropriate to generate evading malware variants. Both attackers and defenders can leverage this information as follows.
On the one hand, attackers know which transformation operators they can discard to speed-up the search for evading malware and minimize calls to the oracle.
On the other hand, defenders can focus on the most effective transformations to evade antiviruses.
For example, one way to defend against evasion is to use  transformation as a preprocessing stage prior to detection. 
This can help to ensure that detection is more robust to potential evasion vectors.

\begin{figure}
  \includegraphics[width=\linewidth]{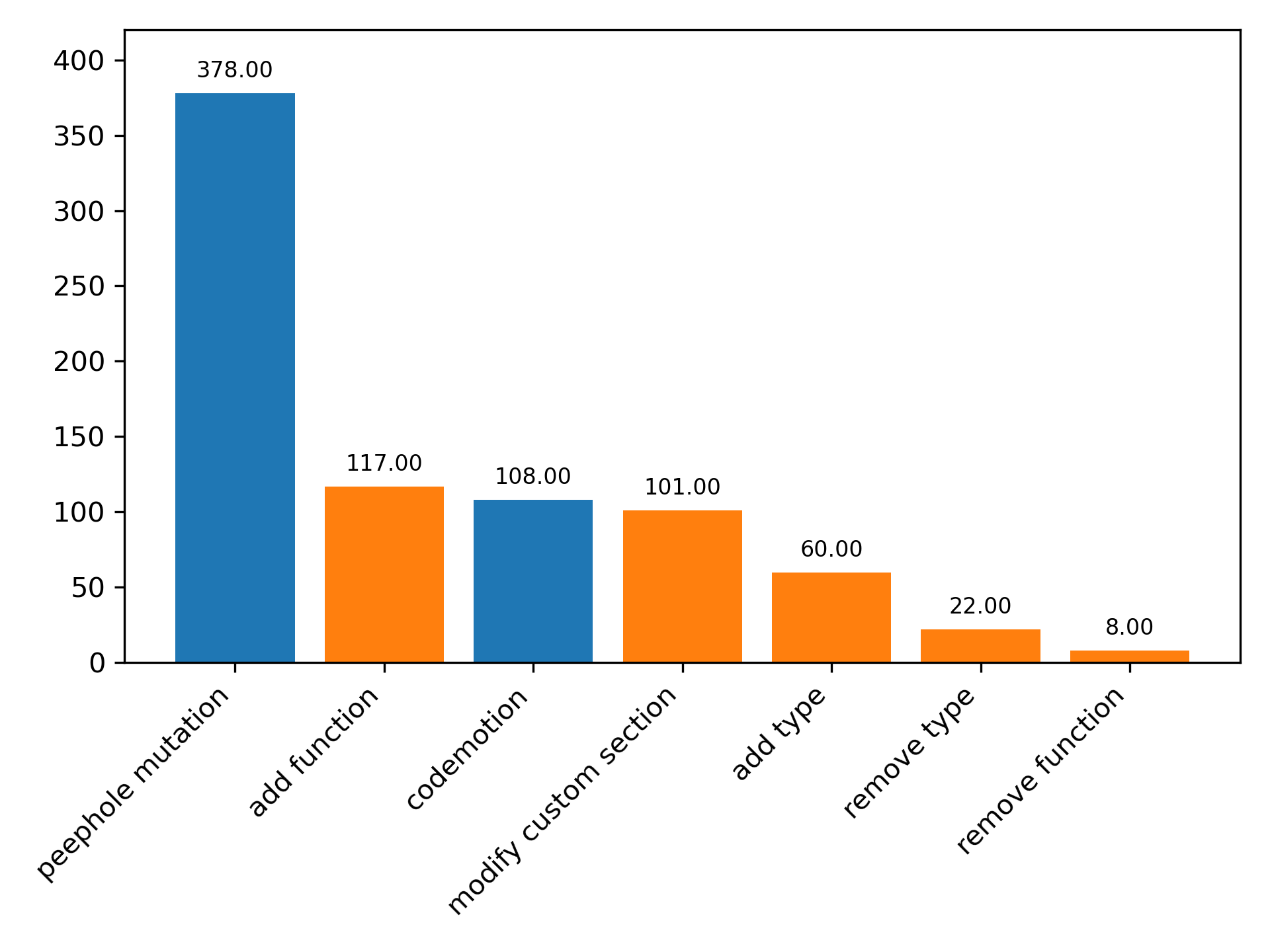}
  \caption{Distribution of applied transformation for the MCMC evasion algorithm with $\sigma=1.1$ (low acceptance).
The x-axis displays the names of the transformations.
The y-axis indicates the number of transformations found in the generated variants.
We can observe which transformations perform better in order to provide total evasion.
}
  \label{plot:profiling}
\end{figure}

In \autoref{plot:profiling} we plot the distribution of transformations applied by the MCMC evasion algorithm with $\sigma=1.1$ when it generates variants that fully evade the VirusTotal oracle.
On the x-axis, we provide the name of the transformation, as wasm-mutate implements them.
The y-axis is the absolute number of transformations found among the generated variants.
The transformations are sorted in decreasing order of usage in the variants.
We use two colors for transformations:  transformations that affect the execution of the binary (in blue color) and transformations that do not (in orange color). 
For example, adding a new type definition does not affect the execution of the binary, while a peephole transformation does.

% Sanitization and expected result in modify custom section
% And the presence of bloat

First, we concentrate on non-behavioral transformations in orange.
The most effective transformation is to add a random function, which is present 117 times in the evading binaries.  
The next most present structural transformation is the \texttt{modify custom section} transformation.
This highlights the sensitivity of malware detectors to the custom section.
We note that custom section modification at the bytecode level provides advantages against source code obfuscation because we are sure that the compiler does not add metadata information that would help malware detectors.
In other words, metadata information injected by compilers into WebAssembly binaries \cite{Hilbig2021AnES}, could be removed from being part of possible detection.

%\todo{R1 - - Whereas the results for individual transformations is important, it is more likely that authors will use transformations in conjunction. So, I suggest authors to consider tuples of transformations in these analyses.}

Transformations \texttt{remove function} and \texttt{remove type} also affect malware detection. 
This novel observation indicates that VirusTotal is looking for code common in malware yet dead code.
For example, malware detectors probably check for code that does not affect the original functionality of the binary. Thus, if this information is removed, the detector misses the malware. In other terms, by removing code, the detection surface is reduced.

% And the best transformation award goes to...
The most significant transformation to generate evading malware consists of peephole transformations (the largest bar of the figure at the left-hand side of the figure).
The peephole transformations operate at the bytecode instruction level.
For example, in \autoref{example:block1} and \autoref{example:block2} we show a piece of the binary \texttt{089a3645c}, and one peephole transformation applied, respectively.
The transformation in the listings corresponds to a variant generated with the MCMC evasion algorithm.
Only this transformation is required to fully evade for the \texttt{089a3645c} binary.

% main white hat takeaway
Our results show that malware detectors should prioritize the detection of peephole transformations in WebAssembly, to increase the likelihood of detecting cryptojacking.
For example, the transformation of \autoref{example:block2} can be reversed with static analysis.

\lstdefinestyle{watcode}{
  numbers=none,
  stepnumber=1,
  numbersep=10pt,
  tabsize=4,
  showspaces=false,
  breaklines=true, 
  showstringspaces=false,
    moredelim=**[is][{\btHL[fill=black!10]}]{`}{`},
    moredelim=**[is][{\btHL[fill=celadon!40]}]{!}{!}
}

{\captionsetup{width=0.45\linewidth}
\noindent\begin{minipage}[b]{0.45\linewidth}
    \lstset{
        language=WAT,
        style=watcode,
        basicstyle=\footnotesize\ttfamily,
        columns=fullflexible,
        breaklines=true}
        \begin{lstlisting}[label=example:block1,caption={Original piece of code for the \texttt{089a3645c} WebAssembly binary.},frame=b, captionpos=b]{Name}
...
local.get 0




i32.const 4
i32.add
...
        \end{lstlisting}
        \vspace{0.26cm}
   \end{minipage}\hfill%
\noindent\begin{minipage}[b]{0.45\linewidth}
    \lstset{
        language=WAT,
                        style=watcode,
        basicstyle=\footnotesize\ttfamily,
                        columns=fullflexible,
                        breaklines=true}
        
        \begin{lstlisting}[label=example:block2,caption={Peephole mutations of \autoref{example:block1}. Only with this mutation VirusTotal passed from 2 initial detectors to 0 in our experiments.},frame=b, captionpos=b]{Name}
...
local.get 0
!i64.const -461681990785514485!
!drop!
!i32.const 0!
!i32.shr_u!
i32.const 4
i32.add
...
        \end{lstlisting}
\end{minipage}
}

% Cherry on the cake
When we answer our four research questions, we generate WebAssembly cryptominer variants by adding one transformation at a time (See \autoref{sec:evasion-algorithms}).
This method allows us to answer our research questions at a fine-grained level.
For instance, the answer to RQ4 could only be possible if the transformations are analyzed one by one in the evasion process.
Now, our method can be easily tuned to one-shot evasion: the algorithms could apply multiple transformations simultaneously to produce evading malware in one iteration.
Consequently, the evasion process proposed in this work could be faster and more practical for a potential attacker.
On the other hand, our algorithms stop as soon as one binary is diversified enough to provide total evasion.
Since the overhead introduced by wasm-mutate is imperceptible, the transformation process can generate remarkably more binaries.
Our approaches could escalate to infinite cryptojacking variants.

\begin{tcolorbox}[boxrule=1pt,arc=.3em,boxsep=-1.3mm]
  \textbf{Answer to RQ4}: 
  Our experiments reveal that peephole transformations are the most effective for WebAssembly \rev{cryptojacking} malware evasion. 
We also show that transformations on non-executable parts of WebAssembly binaries can contribute to evasion. These novel observations are crucial for \rev{cryptojacking} malware detector vendors to prioritize their work on improving malware detection.
\end{tcolorbox}

\subsection{RQ5. Effectiveness against MINOS}
\label{rq5}

To evaluate the effectiveness of MINOS at detecting diversified malware, we reuse the protocol of  RQ1. 
We repeatedly stack random mutations to the original malware binary until MINOS is fully evaded or the maximum number of iterations is reached. We repeat this process 10 times for each binary.
The results of our experiment are presented in \autoref{tab:minos}. The table provides the identifier of the program and the mean number of iterations required to synthesize a variant that fully evades MINOS. 

Our technique completely evades MINOS in all cases.
%Why? Low resolution ? Apparently this is not correlated with the size of the binary
In 2 cases out of 33, wasm-mutate needs more than 900 iterations to evade MINOS.
The main reason is the application of symmetric mutations.
For example, in some cases, wasm-mutate performs mutations that copy parts of the binary to another program location.
Thus, when the binary is turned into a grayscale image, the embedding of the image is preserved, i.e., the code has changed, but the shape of the image has not.
The contrary happens when non-symetric mutations are applied. 
For example, removing functions also removes embeddings of the grayscale image used by MINOS. 

% Now what happen when we pass then to VT ? 

Remarkably, this experiment shows that WebAssembly diversification requires fewer iterations to  evade MINOS than VirusTotal, meaning that it is easier to evade MINOS.
The minimum number of iterations needed overall for evading VirusTotal are 118 for the baseline algorithm \autoref{tab:evasion_t1} and 11 for the MCMC algorithm \autoref{tab:evasion_t2}, while for MINOS, wasm-mutate totally evades detection for 8 out of 33 binaries in one single iteration.
This shows that the MINOS model is fragile wrt binary diversification. 
According to those results, VirusTotal can be considered better than MINOS wrt to cryptojacking detection.

To further enhance the detection capabilities of MINOS, we believe in binary canonicalization \cite{4140990}. By creating a canonical representation of the malware variant before training and inference, one would help the classifier to better generalize. This is feasible as it is a preprocessing step in the pipeline. We believe this is an interesting direction for future work.

\begin{table}[t]

  \caption{The table provides the identifier of the program as its sha256 hash value and the mean number of iterations required to totally evade MINOS. Each binary was mutated with the baseline algorithm 10 times. }
    \label{tab:minos}
  
\centering
\footnotesize
\renewcommand\arraystretch{1.10}
%\begin{adjustbox}{width=1\linewidth}
    \begin{tabular}{lr | lr }
        \toprule
         Hash &  Mean \#trans.& Hash &  Mean \#trans. \\
         %& & \textbf{hash} &  \textbf{\#D}  &  \textbf{Max. \#evaded} &   & \textbf{hash} &  \textbf{\#D}  &  \textbf{Max. \#evaded} & \\
        \hline
          
             24aae13a & 980.0 & 000415b2 & 960.0 \\
             59955b4c & 38.0  & 119c53eb & 1.0  \\
             fb15929f & 1.0  & 5bc53343 & 33.0  \\
             47d29959 & 100.0  & dc11d82d & 115.0  \\
             a27b45ef & 33.0  & 006b2fb6 & 1.0   \\
             942be4f7 & 29.0  & 7c36f462 & 85.0  \\
             0d996462 & 24.0  & 15b86a25 & 1.0  \\
             8ebf4e44 & 92.0  & a74a7cb8 & 38.0  \\
             fbdd1efa & 1.0  & 089dd312 & 68.0  \\
             65debcbe & 38.0  & aafff587 & 1.0  \\
             046dc081 & 33.0  & 6b8c7899 & 38.0  \\
             a32a6f4b & 1.0  & d2141ff2 & 81.0  \\
             68ca7c0e & 38.0  & dceaf65b & 66.0  \\
             9d30e7f0 & 419.0  & 4cbdbbb1 & 1.0  \\
             643116ff & 47.0  & c1be4071 & 38.0  \\
             e09c32c5 & 15.0  & f0b24409 & 33.0  \\
             89a3645c & 108.0 
    \end{tabular}
%\end{adjustbox}
\end{table}

\begin{tcolorbox}[boxrule=1pt,arc=.3em,boxsep=-1.3mm]
  \textbf{Answer to RQ5}: Our approach  fully evades detection by the WebAssembly antivirus MINOS. In our study, we achieve evasion for all cryptojacking binaries in our dataset. wasm-mutate needs fewer iterations to totally evade MINOS compared to VirusTotal, validating VirusTotal as a baseline.
\end{tcolorbox}

\section{Discussion}
\label{discussion}

%\todo{R1- dd a discussion section. I believe discussion section is the most important of a paper, as it answer the question "Now that we know these results, what should we do?". The authors touched this answer in parts of the paper, such as in RQ4, but in discussion they can go deeper.
%}

In this section, we discuss the key challenges we faced, in order to help future research projects on similar topics.

\emph{Dataset size:} The dataset  is smaller than other similar works for malware evasion.
However, the related work does not consider WebAssembly --  e.g. Ling et al. \cite{ling2023adversarial} focus on Windows.
For example, while Tekiner and colleagues  consider cryptojacking \cite{9566204}, we entirely focus only on WebAssembly cryptojacking malware.
In this context, to the best of our knowledge, wasmbench is the most complete dataset of WebAssembly binaries.
We analyze this dataset through the lens of VirusTotal and systematically extract all the cryptojacking malware it includes.
\rev{Despite novelty, we acknowledge that the limited size of our malware dataset poses a challenge in terms of the generalizability of our results. 
Some types of malware might be absent from our dataset. 
To address this issue, one solution for future work would consist in expanding the dataset by using the inherent diversity found within the popular Cryptonight library.
One could utilize the release history of their GitHub repository to compile and mine distinct, yet semantically equivalent malware instances. 
This approach would entail the exploration of a broader range of variations of cryptojacking malware 
Yet, this is considered as a new research paper per se as the process of mining and compiling code from a repository's release history is both time-consuming and computationally demanding. This is due to the need to analyze, filter, and compare a vast number of code commits and releases, as well as to validate their semantic equivalence. This process is even more complex in the case of malware reproduction due to the complex architecture in which this type of software operates (cf. \autoref{attack-model}.)}

\emph{VirusTotal Observations:}
The final labelling of binaries for VirusTotal vendors is not definitive \cite{251586}. 
For example, a VirusTotal vendor could label a binary as benign and change it later to malign after several weeks.
This phenomenon creates a time window in which slightly changed binaries (fewer iterations in our case) sometives evade the detection of numerous vendors.
Also, we have observed that when our evasion algorithms manage to evade, some VirusTotal vendors result in timeout in several cases. 
This suggests that the evasion effectiveness is also due to performance constraints on the VirusTotal side.

\emph{Lack of abstraction:} A WebAssembly cryptojacking can only exist with its web complements.
As we previously discussed, a browser cryptojacking needs to send the calculated hashes to a cryptocurrency service.
This network communication is outside the WebAssembly accesses and needs to be delegated to a JavaScript code.
% Besides, other functionalities can be intermixing between JavaScript and WebAssembly and in some cases be completely in one side or the other \cite{romano2022wobfuscator}.
% What we saw
We have observed that, the imports and the memory data of the WebAssembly binaries have a high variability in our  dataset.
The imported functions from JavaScript change from binary to binary.
Their data segments also differ in content and length.
This suggests that the whole JavaScript-WebAssembly polyglot package is the right direction for cryptojacking detection.

%\emph{More narrowed fitness function:} In the MCMC algorithm, we use a simple fitness function. 
%Yet, the MCMC evasion algorithm could have a fitness function as general as wanted.
%In our case, we do not use binary metadata, instead we focus on the result from the malware oracle, given that the main goal is to evade it.

\emph{Mitigation: } As we noted in our response to RQ4 and RQ5, we believe that code canonicalization and is a promising mitigation technique if they are applied directly to WebAssembly.
% The most frequently observed type of successful mutation in our experiments was peephole optimization, and we have found that transformations to these peepholes are not only feasible, but also cost-effective with wasm-mutate and the EGG project \cite{egg}.
One way to do this would be to modify a diversifier such as wasm-mutate into a binary optimizer completely based on e-graph.
This would provide canonicalization through a compact representation of WebAssembly code. 
In turn, malware variants with the same ancestor would be more seen as the ``same'' program, from its canonical representation. 
% We stress the importance of carrying out these transformations in a WebAssembly-to-WebAssembly manner.  This is because, as we previously discussed, source-to-source or compiler-based transformations can sometimes lead to increased binary metadata, i.e. including bias into datasets.

%\todo{not clear and convincing. TBD}

%\todo{R3.10: The number of samples used for the experimental validation is minimal. Thirty samples are very limited, while other works [1] use 6k samples for their analysis of crypto-miners.}
%\todo{R1 - The points presented by the authors are interesting, but they are more threats to the reproduction than to the validity. At least in my view, validity is related to generalization. This is exactly the biggest limitation of this work. Authors generalize the correctness of 6 samples to 33 and later results from 33 to a whole class of threats. Authors should acknowledge this limitation. Whereas authors presented the best evaluation possible at the moment, it is important to state that the constructions used by the 33 Wasm malware are likely a very very reduced set of all possible malicious constructions possible in Wasm, thus distinct results might be observed.
%}

\section{Conclusion}
\label{conclusions}

We have demonstrated the potential for WebAssembly cryptojacking malware to be diversified and evade detection by leading malware detectors, such as VirusTotal and MINOS. 
Our generated variants are functional, performant, and do not trigger malware detection.
Our evaluation of the technique against 60 state-of-the-art antiviruses through the meta-tool VirusTotal highlights the superiority of meta-antiviruses over single tailored defenses, even when the latter are specifically designed for WebAssembly cryptojacking binaries such as MINOS.
By studying effective code transformations for evading cryptojacking detection, our work provides valuable insights and guidance for researchers in the field to better mitigate evasion.

As future work, we will improve the evasion fitness functions by including malware program properties, w.r.t. both evasion and malware execution performance.
Some argue that the future of malware detection lies in machine learning. 
In future work, we believe in using our diversification technique to provide data augmentation for better malware detection.
%\vspace{-.2cm}

\bibliographystyle{cas-model2-names}

% Loading bibliography database
\bibliography{main}

\balance

\end{document}